\documentclass[9pt,twocolumn,twoside]{pnas-new}
\templatetype{pnasresearcharticle} % Choose template 
\usepackage{booktabs} % for much better looking tables
\usepackage{multirow}
\usepackage{subfigure}
\usepackage{bbm}
\usepackage{dsfont}
\usepackage{bm}

\newcommand{\matr}[1]{\ensuremath{\pmb{#1}}}

\newcommand\invisiblesection[1]{%
  \refstepcounter{section}%
  \addcontentsline{toc}{section}%
}
\renewcommand{\paragraph}{\section}

\setboolean{displaywatermark}{false}

\author[a]{Przemyslaw A. Grabowicz}
\author[b,c]{Francisco Romero-Ferrero} 
\author[d]{Theo Lins}
\author[e]{Fabr\'icio Benevenuto}
\author[f]{Krishna P. Gummadi} 
\author[b,c]{Gonzalo G. de Polavieja}

\affil[a]{The College of Information and Computer Sciences, University of Massachusetts, 01003, Amherst, USA}
\affil[b]{Cajal Institute, Ave. Doctor Arce 37, 28002 Madrid, Spain}
\affil[c]{Champalimaud Centre for the Unknown, Avenida Bras\'ilia, 1400-038 Lisbon, Portugal}
\affil[d]{Computer Department of Federal University of Ouro Preto, Campus Universit\'ario Morro do Cruzeiro, 35400-000, Ouro Preto, Brazil}
\affil[e]{Computer Science Department of Federal University at Minas Gerais, Av. Ant\^onio Carlos, 6627, Pampulha 31270-010, Belo Horizonte, Brazil}
\affil[f]{The Max Planck Institute for Software Systems, Campus E 1 5, 66123 Saarbrucken, Germany}

\leadauthor{Przemyslaw A. Grabowicz} 

\authorcontributions{Author contributions: P.A.G., F.R.-F., F.B., K.P.G, and G.G.d.P. designed experiments and research; P.A.G. and T.L. implemented experiments; P.A.G, F.R.-F., and T.L. performed research; P.A.G and F.R.-F. analyzed data; P.A.G, F.R.-F., and G.G.d.P. contributed analytic methods; P.A.G, F.R.-F., and G.G.d.P. wrote the paper.}
\authordeclaration{The authors declare no conflicts of interest.}
\correspondingauthor{\textsuperscript{2}To whom correspondence should be addressed. E-mail: pms@mpi-sws.org}


\title{Bayesian Social Influence in the Online Realm}
\begin{abstract}
Our opinions, which things we like or dislike, depend on the opinions of those around us. Nowadays, we are influenced by the opinions of online strangers, expressed in comments and ratings on online platforms. 
Here, we perform novel ``academic A/B testing'' experiments with over 2,500 participants to measure the extent of that influence. In our experiments, the participants watch and evaluate videos on mirror proxies of YouTube and Vimeo. We control the comments and ratings that are shown underneath each of these videos. Our study shows that from 5$\%$ up to 40$\%$ of subjects adopt the majority opinion of strangers expressed in the comments. 
Using Bayes' theorem, we derive a flexible and interpretable family of models of social influence, in which each individual forms posterior opinions stochastically following a logit model. The variants of our mixture model that maximize Akaike information criterion represent two sub-populations, i.e., non-influenceable and influenceable individuals. 
The prior opinions of the non-influenceable individuals are strongly correlated with the external opinions and have low standard error, whereas the prior opinions of influenceable individuals have high standard error and become correlated with the external opinions due to social influence. 
Our findings suggest that opinions are random variables updated via Bayes' rule whose standard deviation is correlated with opinion influenceability. 
Based on these findings, we discuss how to hinder opinion manipulation and misinformation diffusion in the online realm.

\end{abstract}

\dates{This manuscript was compiled on \today}
\doi{\url{www.pnas.org/cgi/doi/10.1073/pnas.XXXXXXXXXX}}

\begin{document}

% Optional adjustment to line up main text (after abstract) of first page with line numbers, when using both lineno and twocolumn options.
% You should only change this length when you've finalised the article contents.
\verticaladjustment{-2pt}

\maketitle
\thispagestyle{firststyle}
\ifthenelse{\boolean{shortarticle}}{\ifthenelse{\boolean{singlecolumn}}{\abscontentformatted}{\abscontent}}{}

% opinion formation - from offline to online
%Our opinions are influenced by the opinions of others \cite{Asch1955Opinions, Sherif1935study, Kelman1961Processes, Wood2000Attitude, Cialdini2004Social, Turner1987Rediscovering, Regier2009Language}. 
In the previous century, experts heavily influenced the public opinion via mass media~\cite{Lippmann1922Public, Herman1988Manufacturing}. Nowadays, billions of individuals express their opinions in the online realm through online comments, reviews, and evaluations~\cite{Richtel2013Theres}. 
Unfortunately, it is relatively easy and cheap to fake such online opinions, compared with physical world~\cite{King2016How, Cho2011Astroturfing}.\footnote{There exist several websites selling comments, thumbs up, and views in social media, for instance \url{https://buysocialmediamarketing.com} and \url{https://www.qqtube.com}. Last checked in April 2018.} 
In the recent years, so-called astroturfers were hired to proliferate selected opinions and fake news online during major societal events, including democratic elections~\cite{Lazer2018science,Vosoughi2018spread}. To design systems that are robust to misinformation and manipulation, it is crucial to uncover the extent and the mechanism of social influence. 
%On the black market one can buy various types of fake social feedback, including comments and positive evaluations~\cite{2016There}.
%This issue is counterbalanced by low quality of such online opinions, so it is not clear whether the opinions expressed online by strangers influence our opinions and whether some types of social feedback are more influential than others. 
%Here, we test whether the opinions of strangers influence public opinion, measure the extent of that influence for different types of social feedback, and propose a flexible probabilistic theory and models describing the mechanism of social influence.

% experimental challenges: external validity, causality
% soultion: academic A/B testing experiments
The challenge in experimental studies of social influence is that traditional lab experiments~\cite{Asch1955Opinions, Sherif1935study} and online surveys~\cite{Eguiluz2015Bayesian} lack external validity. Field experiments, on the other hand, yield less control over confounding factors, hindering causal inference~\cite{Muchnik2013Social, Wang2014Quantifying, Sipos2014Was, Kramer2014Experimental}. 
%We circumvented these experimental issues by designing a hybrid of an online survey and a field experiment. 
To address these experimental challenges, we conduct novel \textit{academic A/B testing} experiments. We create ``clones'' of existing social media websites, i.e., mirror proxies of real websites that are fully controlled by researchers to perform randomized experiments. These mirror proxies have exactly the same look and basic functionalities as their real counterparts.
% our experiments
In our experiments, the participants watch and evaluate videos on YouTube and Vimeo, i.e., popular video-sharing platforms, in their private spaces and comfort zones. 
Underneath each of the videos, we show different types of social feedback, including the comments and the counters for views, likes, and dislikes. In our experimental conditions, we randomly modify this social feedback by suppressing some of the comments and lowering the counters. Then, we survey the participants about their opinions on each of the videos, to find whether the social feedback they are exposed to influences public opinion.
% awareness
The participants are unaware that this social feedback is modified, but they are informed that the goal of the experiments is to survey their opinions.
Among others, we quantify the extent of social influence of online strangers on the opinions of participants and test whether the comments or the counters exert more influence.
Thanks to our novel A/B testing setup, these measurements are precise and yield high external validity.

The results of social influence experiments are typically explained with the socio-psychological theories of \textit{informative} and \textit{normative} social influence \cite{Kelman1961Processes, Wood2000Attitude, Cialdini2004Social}. The former theory states that social influence stems from our need for accurate information about real world and that this influence is facilitated by the objective perceptual uncertainty about the stimulus. 
The theory of normative influence explains social impact with the need to be accepted by others, arising when a mutual relationship between individuals is present. 
However, social influence is observed even if these conditions are not met \cite{Turner1987Rediscovering}. Indeed, the participants of our experiments watch and evaluate videos that do not exhibit objective perceptual uncertainty and they are exposed to social signals from online strangers. 
The more recent theory of \textit{self-categorization} addresses these points and explains the results of seminal experiments on social influence~\cite{Sherif1935study,Asch1955Opinions} by introducing subjective uncertainty and arguing that social signals interfere with that uncertainty in a way that informative and normative needs are inseparable \cite{Turner1987Rediscovering}. 
%This theory posits that individuals influence each other if they self-categorize themselves with a common identity, e.g., a stratum of social media users.
%The theory postulates that the agreement between individuals sharing common identities validates socially their opinions, whereas disagreement between them creates subjective uncertainty about their judgments.  
%The self-categorization theory posits that social influence is facilitated by the uncertainty due to objective perceptual ambiguity as well as subjective uncertainty and broadens the sense of the latter to uncertainty caused by social disagreement with individuals sharing a common identity. 
%This theory posits that social influence is facilitated by the uncertainty due to objective perceptual ambiguity as well as subjective uncertainty. Here, we study this relationship between influenceability and uncertainty based on probabilistic reasoning.
% probabilistic reasoning
Human judgments under uncertainty have been extensively studied in psychology~\cite{Phillips1966Conservatism, Tversky1974Judgment, Navajas2017idiosyncratic, Benjamin2019Errors}. Pioneering works in this area measure human biases in probabilistic reasoning by means of comparisons with Bayesian inference, in particular the binomial model~\cite{Edwards1963Bayesian, Phillips1966Conservatism, Grether1980Bayes, Benjamin2019Errors}. 
Here, we derive the corresponding binomial model of social influence from basic principles of probability theory, following empirical Bayes method. 
In our settings, priors and posterior distributions are unknown and estimated from observations.
%, whereas in the context of judgments under uncertainty the posterior probabilities are unknown~\cite{Benjamin2019Errors}.
%This model fits the data well and captures the relation between influenceability and uncertainty, but it does not model it directly.
%With this model and its generalization, 
With this model, we measure social influenceability and its relation to uncertainty in online social media and introduce a Bayesian theory of social influence.

%
%the risk, concerning ``known unknowns'', but also by uncertainty, corresponding to ``unknown unknowns''\cite{}. 
%Here, we confirm this implication experimentally using the mathematical formalism of a Bayesian theory of social influence. Our study suggests that there is a parallel between self-categorization theory and the Bayesian theory of social influence.
%Here, we study the relation between influenceability and uncertainty experimentally using the mathematical formalism of a Bayesian theory of social influence, which is based on the same basic assumption of social validation, as the self-categorization theory. 
%Thus, we suggest that there is a connection between social psychology and Bayesian statistics.
%The uncertainty and confidence of human decisions can be estimated with the observed Fisher information~\cite{Navajas2017idiosyncratic}. We measure the observed Fisher information of opinion, to test the hypothesis that social influence is facilitated by the uncertainty in opinion.

%Here, we explain our experimental results with a Bayesian theory of social influence, which collapses the informative and normative social influence into one as well. 
%We identify groups of individuals by their reaction to the social feedback and inquire whether they comply with these social signals.

\section*{Results}

\begin{table}[t]
\centering
\caption{Summary of the experiments.}
\begin{tabular}{lrr}
							& Experiment I 		& Experiment II		\\
\toprule
Platforms:					& YouTube, Vimeo	& YouTube			\\
Participants:				& 1,116 				& 1,391				\\
Videos:						& 8					& 14				\\
Experimental conditions:	& 11				& 7					\\
Expressed opinions:				& 8,928				& 9,737				\\
Opinion scale:				& 5-point Likert	& 200-point bipolar	\\
Comment tracking:	& No		& Yes	\\
\bottomrule
\end{tabular}
\label{tab:confusion}
\end{table}

% size
In total, over 2,500 subjects participated in our experiments (Table~\ref{tab:confusion}).
% description
The participants were recruited and compensated via Amazon Mechanical Turk~\cite{Buhrmester2011Amazons}.
Each participant of our online experiments is instructed, in one sitting, to watch several videos, familiarize themself with social feedback to these videos, and answer whether they like the video. 
% real platforms and content
Each web page, showing a video with corresponding user comments, is a clone of an existing web page on YouTube or Vimeo.
The videos are on a variety of topics, ranging from pranks and commercial ads to societal issues and innovations.
We selected videos that had up to a million views at the moment of data gathering, but not more, to avoid potential confounding effects from participants who have seen them before.
% reception
Albeit the videos were pre-selected, many participants of our experiments found them to be entertaining.
Over 185 participants used words ``great'', ``fun'', or ``enjoy'' in reference to the videos and the experiment in an optional text-box comment presented at the end of each experiment (see Demographics and Feedback of Participants in SI Appendix).
%More than 185 participants voluntarily expressed the experiments to be ``fun'', ``enjoyable'', and ``great''.
%\footnote{More than 185 participants voluntarily expressed this sentiment at the end of experiment. See Demographics and Feedback of Participants in SI Appendix.}
% participants' recruitment and actions 

\begin{figure}[t]
\centering
\includegraphics[width=0.5\textwidth]{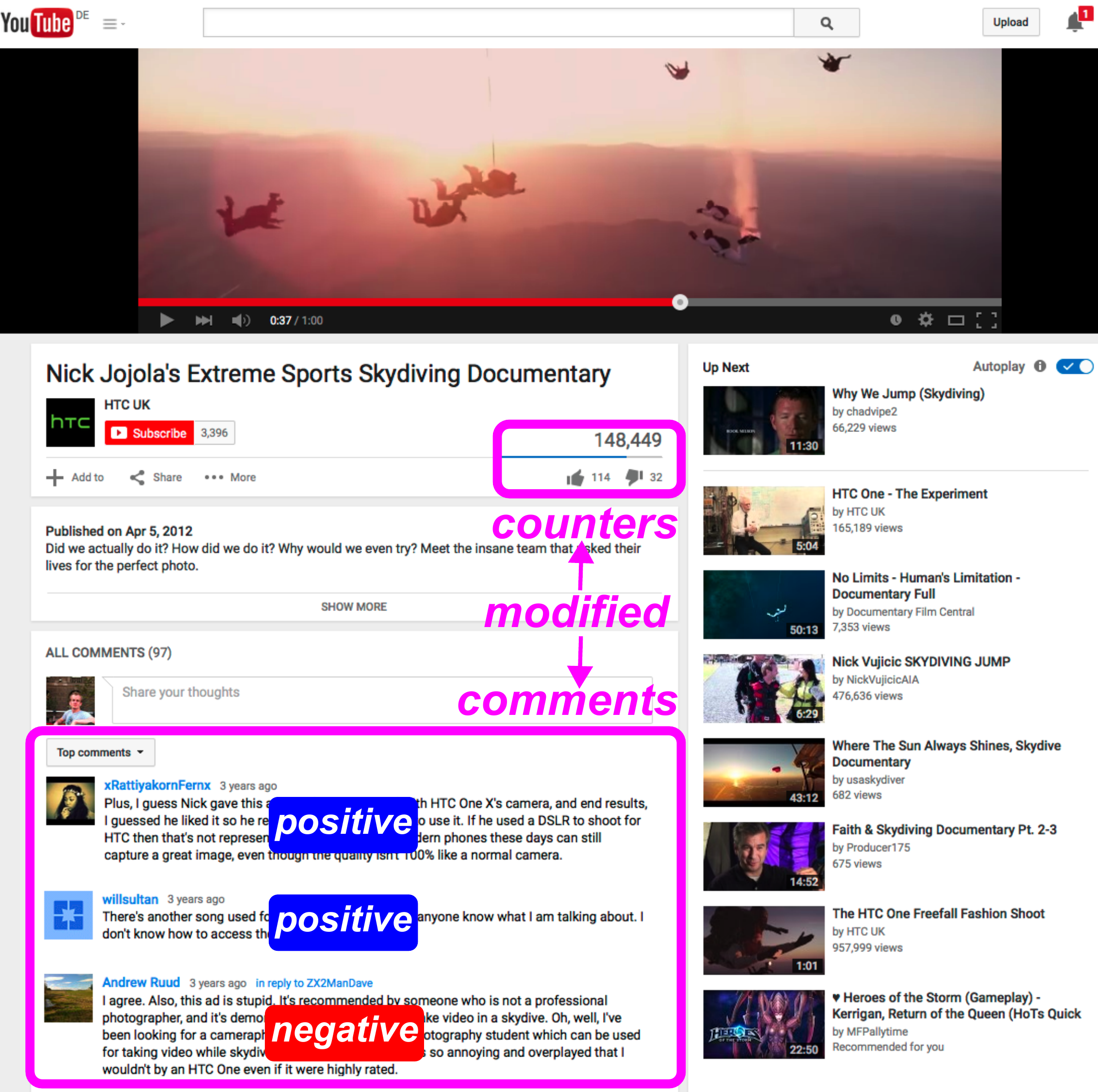}
\caption{An illustration of the experiments. The comments are labeled as positive or negative towards the video before the experiments. Under experimental conditions, some of the comments are randomly suppressed and the counters for views and for thumbs up and down are modified.}
\label{fig:yt}
\end{figure}

% social feedback
Our goal is to measure how social feedback influences opinion.
Each video comes with two types of social feedback: i) the comments of its prior viewers and ii) the counters for views, thumbs up, and thumbs down (see Figure~\ref{fig:yt}). 
% experimental conditions
In the experimental conditions, we control and modify these two types of social feedback. 
In the negative experimental conditions, we hide positive comments and lower the numbers of views and thumbs up; whereas in the positive conditions, we make the modifications that are exactly opposite.
Overall, in the experiments there are three \textit{main} negative conditions and three \textit{main} positive conditions, differing in the degree of modifications to social feedback.
% control condition
As the control condition we take the respective video with its original unmodified comments and the values of counters.

%\begin{figure}[t]
%\centering
%\includegraphics[width=0.35\textwidth]{figs2/expmix_survey3-a_treat-correct_resp-avg-opinion-grad_bothsolo}
%\caption{The probability of a positive opinion under the control condition (circle) and the experimental conditions of various strength (squares and stars). The color of markers corresponds to positive (blue) and negative (red) conditions. This result is averaged over all videos and experiments. The $95\%$ confidence intervals are from BCA bootstrap.}
%\label{fig:grad}
%\end{figure}

%\begin{figure}[t]
%\centering
%\includegraphics[width=0.35\textwidth]{figs2/survey2-a_treat-correct_resp-avg-opinion-part_bothsolo}
%%\includegraphics[width=0.35\textwidth]{figs2/survey2-a_treat-correct_resp-avg-share-part_bothsolo}
%\caption{The probability of a positive opinion for the control condition (circle) and the experimental conditions in which either comments or counters (diamonds) or both (stars) are modified.}
%\label{fig:part}
%\end{figure}

\begin{figure*}[t]
\centering
\begin{picture}(400,200)
\put(0,0){
\includegraphics[width=0.35\textwidth]{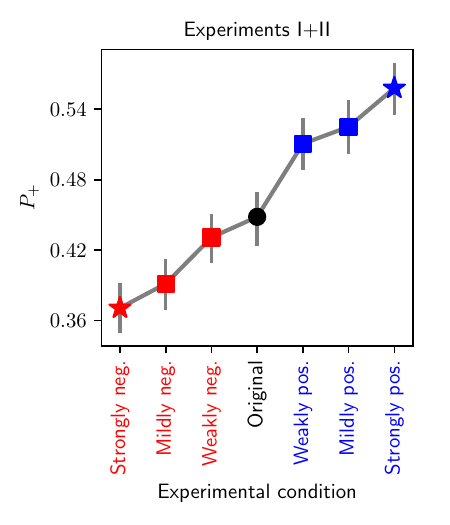}
\includegraphics[width=0.35\textwidth]{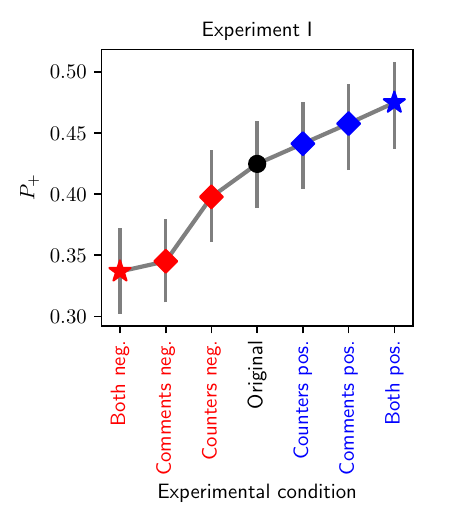}
}
\put(50,175){a}\put(230,175){b}
\end{picture}
\caption{
The probability of a positive opinion under the control condition (circle) and the experimental conditions of various strength (squares and stars). The color of markers corresponds to positive (blue) and negative (red) conditions. This result is averaged over all videos and experiments. The $95\%$ confidence intervals are from BCA bootstrap. (A) Main experimental condition. (B) Partial experimental conditions in which either comments or counters (diamonds) or both (stars) are modified.
}
\label{fig:grad}
\label{fig:part}
\end{figure*}

% the probability
Then, we compute the probability of positive opinion about any video, $P_+$, as the fraction of positive answers across users and videos in the given experimental condition. 
% figures for main experimental conditions
This probability increases monotonously with the extent of modification of social feedback, ordered from the most negative to the most positive main experimental condition (Figure~\ref{fig:grad}A).
In other words, the more positive comments, thumbs up, and views a participant sees under a video, the more likely they are to have a positive opinion about that video.
% confidence intervals
The $95\%$ confidence intervals show that the difference in the probability of positive opinion between experimental conditions is statistically significant for most of the condition pairs.
% statistical tests for differences between distributions
Comparisons of the distributions of raw responses confirm this result. 
For instance, there is a statistically significant difference in opinions between the control condition and the strongly positive and negative conditions (Mann-Whitney U test, $p<10^{-17}$ and $p<10^{-4}$, respectively). 
We conclude that the opinions are influenced by social feedback both positively and negatively~\cite{Kramer2014Experimental}. 
%Therefore, we find the evidence of opinion change.
%\footnote{

% the two experiments
%Here, we show the combined results for both Experiment I and Experiment II, but we also obtain nearly the same results if the analysis is performed separately for each of the experiments.
% information diffusion and extras
In addition to opinion, we also survey the participants about their willingness to share the video they watched with friends. The willingness to share a video is correlated with the opinion about that video, because positive opinion about an object creates incentives for sharing it with friends~\cite{Heider1958Psychology}. We find that all presented results are qualitatively and often quantitatively the same for the opinion and the sharing willingness, under different psychometric scales (see Experiment I and Experiment II in SI Appendix).
%}

% partial experimental conditions
So far we have shown the result for the main experimental conditions, in which both the comments and the counters are modified. However, it is not clear whether the participants are influenced by the comments or the counters. To answer this question, in Experiment I, we measure which type of social feedback exerts more influence on the opinions: the comments or the counters? To this end, additional experimental conditions are introduced, in which either only comments or only counters are modified, i.e., two positive and two negative \textit{partial} experimental conditions.
%These experimental conditions allow us to measure whether the comments or the counters influence the opinions more. 

% recalling partial conditions
In contrast to main positive and negative conditions, the partial conditions modify only one type of social feedback, instead of both of them.
%, i.e., they modify only comments or only counters and leave the other type of social feedback unchanged.
% analysis
We find that the probability of positive opinion is influenced more by the modifications of comments than the counters (compare the diamonds of the same color in Figure~\ref{fig:part}B).
In other words, the comments have larger impact on opinion than the thumbs and views. The comments significantly influence the opinion in negative and positive experimental conditions with respect to the original condition ($p=0.0002$ and $p=0.02$, respectively), whereas the influence of thumbs is insignificant.

\begin{figure*}[t]
\centering
\includegraphics[width=0.90\textwidth]{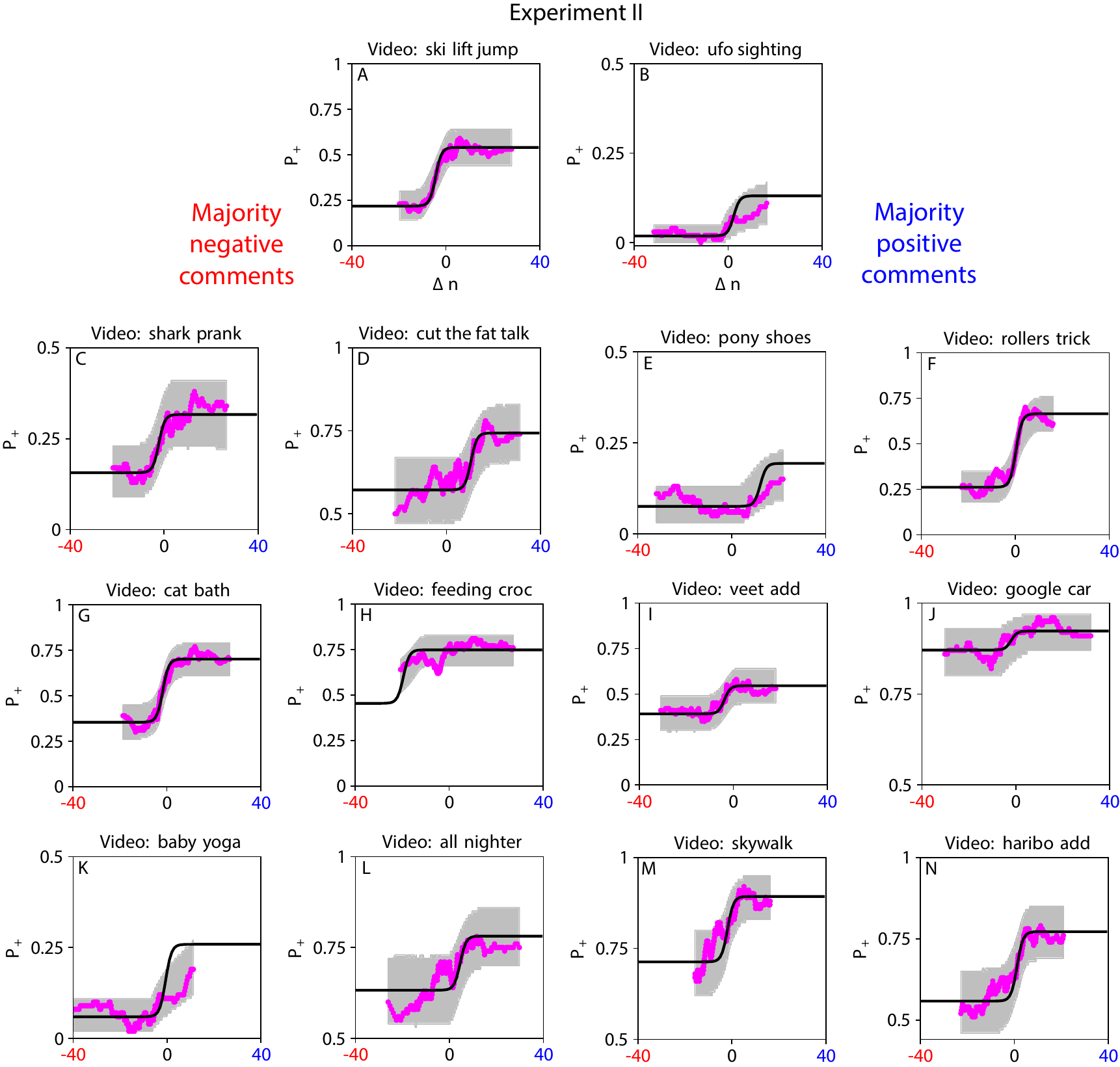}
\caption{Public opinion about a video depends on majority opinion in comments. 
The probability of positive opinion is plotted versus the difference in the number of positive and negative comments read by a subject. Magenta points correspond to the moving average over $100$ measurements. The model applied to the finite real data expects the running average in the gray area marking $99\%$ confidence interval. The black line is the model applied to an infinite data.
}
\label{fig:vids}
\end{figure*}

% versus the valence of consumed comments
In the remainder, we explain the mechanism of social influence by analyzing in more detail the influence of comments on public opinion.
In Experiment II, to better understand the mechanism of social influence, we make more precise measurements for more videos and participants, while tracking which comments were read by each of the subjects.\footnote{We track which exact comments are shown on the screen of each participant.}
%In Experiment II, we track which comments each participant reads.\footnote{We track which exact comments are shown on the screen of each participant.}
We exploit this information to measure how the probability of a positive opinion about a video, $P_{+}$, depends on the difference, $\Delta n$, in the number of positive and negative comments read by a participant (Figure~\ref{fig:vids}). Subjects have a relatively positive opinion about a given video when positive comments prevail among the comments that they read ($\Delta n >> 0$), and a relatively negative opinion if negative comments prevail ($\Delta n << 0$). For each of the videos, the probability of positive opinion $P_{+}(\Delta n)$ saturates at a lower value when $\Delta n$ is very negative and at a higher value when $\Delta n$ is very positive. 
For most videos, the probability $P_{+}(\Delta n)$ has a sigmoid shape and is anti-symmetric, exposing a systematic dependence. 
%On average, the difference between the lower and higher levels is $\Delta P_{+} = 0.17$. 
Next, we use a Bayesian theory and models of social influence, to explain these experimental results and to estimate the percentage of individuals influenced by the comments.\footnote{We release our dataset to scientific community at \url{www.linktodata}.}

%This difference between levels roughly corresponds to a difference of one star on a five-point star rating popular in online websites (see ``Preliminary experiments'' in the supplementary materials).

%With this approach, we derive the probability of positive opinion as the function of the number of positive and negative opinions received from social sources.
%Until now, Bayesian models have explained social influence only for homogeneous individuals of certain animal species, such as fishes and ants~\cite{Arganda2012common,Perez-Escudero2011Collective,Eguiluz2015Bayesian}. 
%However, Bayesian approach is among the most promising for understanding cognitive processes~\cite{Tenenbaum2011How}. 

%\section*{A Bayesian Theory of Social Influence} 

% experimental challenges: accuracy, fundamental?
Prior works proposed models of social influence that are accurate at predicting opinions in specific circumstances~\cite{Wang2014Quantifying, Sipos2014Was, Kramer2014Experimental}. In this study, we explain the mechanism of social influence with a generic Bayesian theory.
% linking statistical and sociological theories
This theory posits that opinions are hypotheses whose probabilities of being correct are updated in the process of social influence, following Bayes' theorem. 
To this end, we assume that social signals serve as evidence validating corresponding opinions. This assumption is equivalent to the basic assumption of the self-categorization theory that agreement with others ``subjectively validates our responses [i.e., opinions] as veridical reflections of the external world''~\cite{Turner1987Rediscovering}.
To make a direct correspondence to Bayesian estimation, we distinguish between \textit{prior} and \textit{posterior} opinions, i.e., latent parameters representing the opinions before and after social influence, respectively.
In particular, using this Bayesian theory, we derive the posterior probability of an individual to express a positive opinion, $P_+$, as a simple logit model, which accounts for the prior opinion of the individual and the opinions expressed by others~\cite{Perez-Escudero2011Collective,Arganda2012common}. 
On the grounds of this generic theory, other models could be proposed as well, but the logit model is particularly simple and sufficiently expressive. The results of our Experiment II are explained by a mixture of logit models, corresponding to the mixture of participants. We explore many variants of this mixture model, which differ in the number of components, using Akaike information criterion. The best models reveal interesting common properties.
%Our analysis of the models that best explain our data yield interesting 
%Our analysis suggests that the prior opinion of individuals is also a random variable, whose standard deviation is correlated with influenceability. 

%\subsection*{Logit Model}

% derivation
%To derive this family of models of social influence, we assume that individuals form opinions following Bayes rule, while accounting for their prior opinions and the opinions of others~\cite{Arganda2012common}. 
The logit model is derived as a result of Bayesian estimation under social feedback only when, instead of purely rational decisions \cite{Bikhchandani1992Theory, Easley2012Networks}, individuals respond with an additional stochastic component known as probability matching, as shown across animal species, including humans \cite{Arganda2012common,Perez-Escudero2011Collective,Eguiluz2015Bayesian,Madirolas2015Improving} (see Derivation of Logit Model in SI Appendix).
Under this model, the posterior probability of a positive opinion is $P_{+}(\Delta n, s, a)= 1/\left( 1+\exp(-s \Delta n - a) \right)$.
%\begin{equation}
%P_{+}(\Delta n, s, a)= \frac{1}{ 1+\exp(-s \Delta n - a) }.
%%1/\left( 1+\exp(-s \Delta n - a) \right)
%\end{equation}
Here, $a$ is the latent parameter representing the prior opinion of an individual, which corresponds to personal memories and thoughts on a given subject, and~$s$ is their influenceability, which describes how strongly influenced they are by the social feedback on that subject. 
%However, there is a crucial difference between this model and our data. 
This model predicts that the posterior probability of expressing a positive opinion in a homogenous population, where all individuals have the same value of parameters $a$ and~$s$, tends to saturate at $0$ and $1$ for $\Delta n << 0$ and $\Delta n >> 0$, respectively. This phenomena is not observed in our experimental data, likely because each individual reacts differently to social influence on a given topic. To include this heterogeneity in the model, we treat the overall population as a mixture of heterogeneous individuals, that is, a mixture of logit models. 
%To this end, we group individuals into sub-populations. 
%A sub-population $k$ is characterized by its prior opinion $a_k$, influenceability $s_k$, and the fraction $p_k$ of individuals that belong to it.
%Next, we compare specific models from this family using Akaike information criterion and we analyze their properties. 
%\subsection*{Sub-population Models}
% model definition
%Then, we search for a mixture of logit models that best explains our data. 
Each component of the mixture correspond to a sub-population of individuals.
%We introduce a macroscopic models of social influence, by assuming that there exist sub-populations of individuals with shared influenceability and/or opinions. 
A~sub-population $k$ is characterized by the fraction $p_k$ of individuals belonging to it, their prior opinion $a_k$, and their influenceability $s_k$. 
% model structure
However, we do not know how many sub-populations there exist and whether their parameters differ between videos or are necessary for explaining the data. 
%For instance, a sub-population of individuals may have different prior opinions on different videos.
To answer these questions, we explore thousands of variants of the sub-population model, differing in the number of sub-populations and parameters.
First, we consider variants having from one ($K=1$) to six ($K=6$) sub-populations.
Second, each of the parameters, $a_k$,$s_k$, and $p_k$, either depends on the video, is constant across videos, or vanishes due to replacement by a neutral constant. 
%In total, for a given number of sub-populations, $K$, there are $N \times \left[ \binom{N^2+K-1}{K} - \binom{N^2+K-3}{K-2} \right]$ unique models of this kind, differing in the combination of parameters, where $N$ is the number of possible implementations of each parameter, namely, i.e., $N=3$. 
% the number of sub-populations
We fit each unique variant of this model to the data by maximizing the likelihood of our observations. To obtain the model that best explains our data, we rank these models by Akaike information criterion~\cite{Stone1977Asymptotic}. 

% charateristics of best models
We then analyze what the best models share in common. The top four models have two sub-populations: a non-influenceable sub-population with $s_1=0$ and an influenceable sub-population with the influenceability $s_2>0$, both of which are constant across videos (see Model Selection in SI Appendix). The top model fits the data remarkably well  (Figure~\ref{fig:vids}). 
% parameters
It has $s_{2}=0.8 \pm 0.005$ across videos, whereas the other parameters, namely $p_1$, $a_1$, and $a_2$, depend on videos. The fraction of influenced individuals varies from $p_2=0.05 \pm 0.05$ to $p_2=0.40 \pm 0.05$, depending on the video (see The Parameters of the Best Model in SI Appendix), which means that a large portion of individuals is influenced by the comments they read.  
%Note that the private opinion of the non-influenciable sub-population, $a_2$, depends less on video, as it is the case in $9$ out of $10$ of the top models. 
% validation
Possibly, the participants are influenced by the comments, because they agree with them. To test this hypothesis, we measure whether the membership in the influenceable sub-population is related to the agreement with comments, self-reported by each participant for each video. 
%To test the consistency of this model, we measure whether the membership in the influenceable sub-population is related to the agreement with comments, self-reported by each participant for each video. 
The membership of a sample in a sub-population generally depends on the likelihood that this sample was generated by that sub-population, i.e., $P_{+}(\Delta n, s_k, a_k)$. This likelihood is significantly correlated with the agreement with comments (Spearman's $\rho = 0.41$, $p<10^{-99}$).
% conclusion
We conclude that social feedback tends to influence the opinion of subjects who agree with the exposed opinions, although it can arise without a conscious agreement as well. 
%We obtain this result without optimizing for it explicitly, so our model and its interpretation are consistent.

\begin{figure}[tb]
\centering
\includegraphics[width=0.49\textwidth]{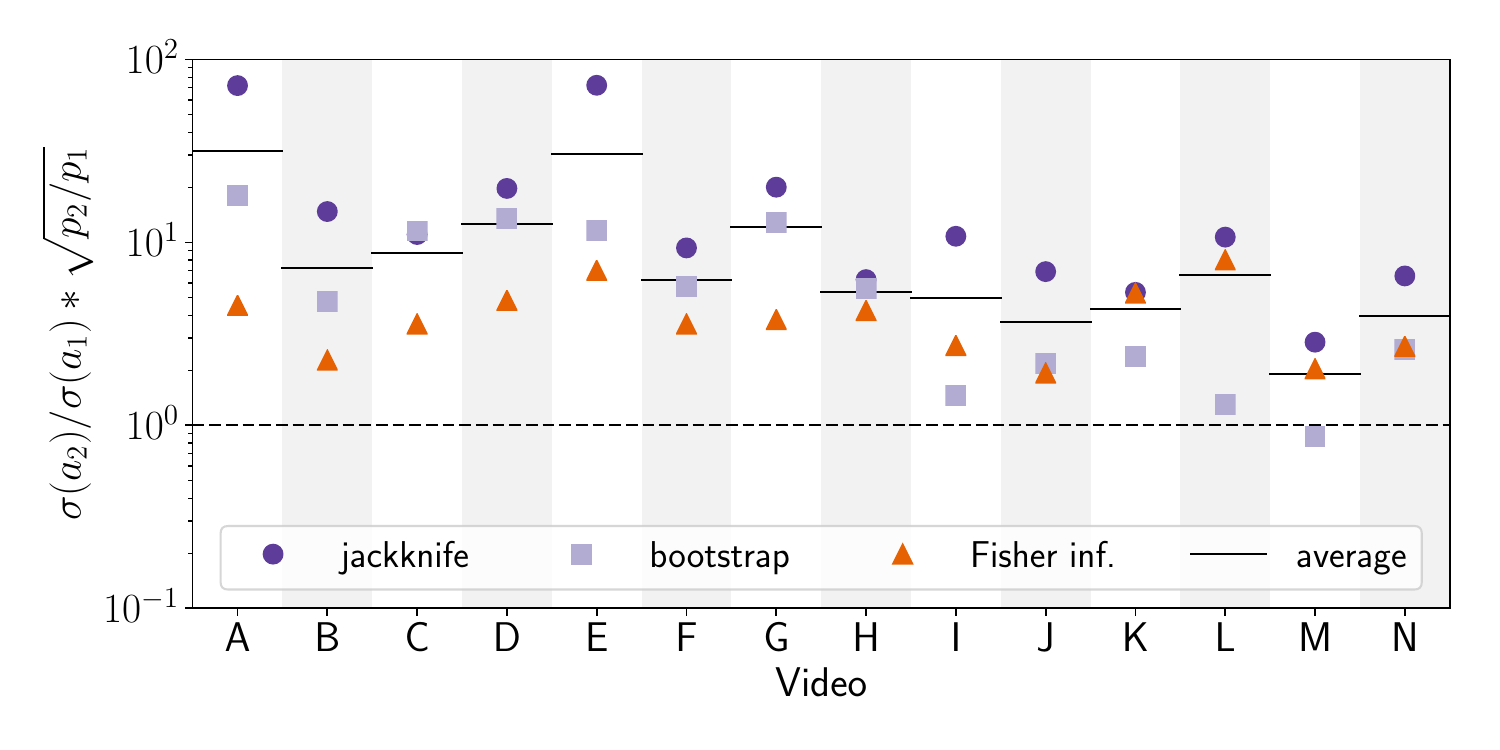}
\caption{The ratio of standard error of prior opinion of the influenceable and non-influenceable individuals. Each black horizontal line is an average over three methods of estimating standard error: as the inverse of observed Fisher information (triangles) or via jackknife resampling (circles) or bootstrap resampling (squares).}
\label{fig:std}
\end{figure}

\begin{figure}[tb]
\centering
\includegraphics[width=0.49\textwidth]{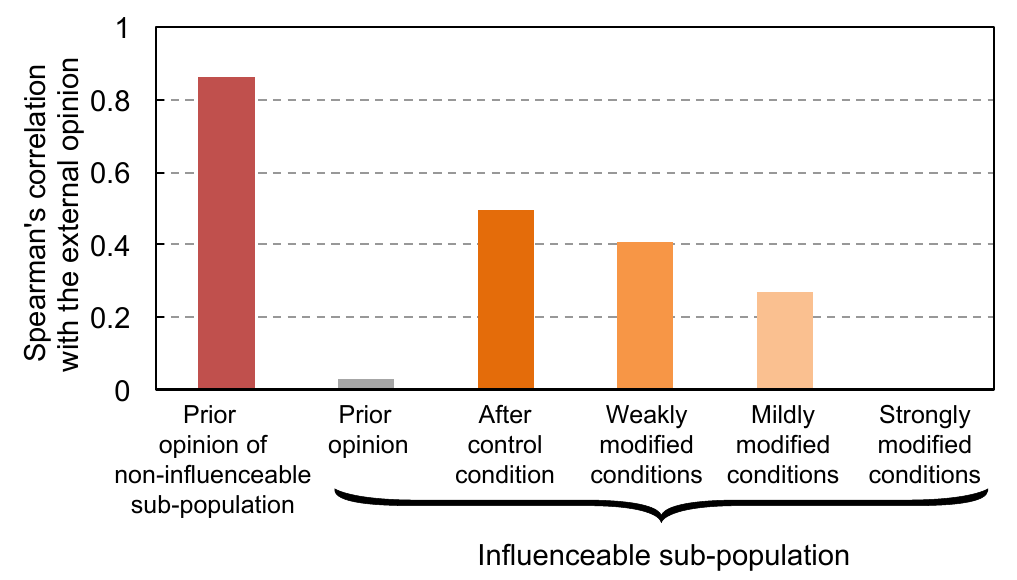}
\caption{The average correlation between the external opinion about videos and the prior and posterior opinions of either non-influenceable or influenceable individuals. As the external opinion we take the fraction of thumbs up for the original video (reported in Table S2 in SI Appendix). As the socially influenced, i.e., posterior, opinions we take the prior opinion on videos plus the effect of social influence, i.e., $a_v + s_2 \Delta n$, where $\Delta n$ is a sample for a random individual for the given video in the experimental conditions of the given kind. The correlation is computed over videos and averaged over the samples.}
\label{fig:corrs-vids}
\end{figure}

% variance of prior opinions
Apart from the difference in influenceability, there are other notable distinctions between the two sub-populations. First, the prior opinion of the influenceable sub-population has a significantly larger standard error in comparison with the non-influenceable sub-population. After correcting for the difference in sizes of the two sub-populations, with the factor $\sqrt{p_2/p_1}$, we find that the standard error of prior opinion of an influenceable individual is from $2$ to $31$ times larger than of a non-influenceable individual (depending on the video, as shown in Figure~\ref{fig:std})A. 
Prior work shows that the standard error of a perceived variable is closely and inversely related to the confidence of human decisions depending on that variable~\cite{Navajas2017idiosyncratic}.
%The uncertainty and confidence of human decisions can be estimated with the observed Fisher information~\cite{Navajas2017idiosyncratic}. We measure the observed Fisher information of opinion, to test the hypothesis that social influence is facilitated by the uncertainty in opinion.
Thus, our result suggests that influenceable individuals are more uncertain in their prior opinions than non-influenceable individuals, making them more influenceable.
This result is in line with the expectations of the self-categorization theory.
%Similarly, the prior opinions of influenceable sub-population do not depend on videos in most top ten models, whereas the prior opinions of non-influenceable sub-population depend on videos. 
%This finding in line with the self-categorization theory and the Bayesian theory of social influence, as we discuss in the next paragraph.
% vulnerability
Second, we find that the prior opinions of non-influenceable sub-population, $a_1$, are strongly correlated with the external opinions ($p<10^{-4}$, the leftmost bar in Figure~\ref{fig:corrs-vids})B. In contrast, the prior opinions of influenceable sub-population, $a_2$, are not correlated with the external opinions (the gray bar in Figure~\ref{fig:corrs-vids})B and become correlated only after being socially influenced (the two middle bars in Figure~\ref{fig:corrs-vids})B.
%\footnote{Here, as the socially influenced opinions we take the prior opinion on videos plus the effect of social influence, i.e., $a_v + s_2 \Delta n$, where as $\Delta n$ we take the random sample for the given video in the experimental conditions of the given kind. The correlation is averaged over these samples.} 
% interpretation
%For these reasons, we refer to prior opinion with large standard error as \textit{weak} opinion, whereas with small standard error as \textit{strong} opinion. 
In other words, social influence helps to develop weak opinions with high standard error into stronger opinions with lower standard error that are closer to the external opinions.
However, under strongly modified experimental conditions, the posterior opinions of influenceable sub-population become heavily distorted and further from the external opinions than their prior opinions (the rightmost bar in Figure~\ref{fig:corrs-vids})B. This result shows that the influenceable individuals are vulnerable to opinion manipulation.

\section*{Discussion}

% explanation - bayes
Our findings provide empirical support for the probabilistic nature of opinion formation in three different ways.
First, in the Bayesian theory of social influence, the prior opinion is updated due to social feedback via Bayes' rule, giving the posterior distribution of opinion~\cite{Bikhchandani1992Theory,Easley2012Networks}.  
The view of social influence, as a social validation of opinions, is consistent with self-categorization theory. Additionally, the theory of self-categorization states that the influence is greater among individuals who share salient identities. The question of whether there is a natural counterpart of this mechanism in Bayesian statistics is open for future research.
Second, the logit model explains our experimental observations only if we apply as the decision rule so-called probability matching, which means that the expressed opinions are drawn from the posterior distribution of opinion~\cite{Perez-Escudero2011Collective,Arganda2012common}.
Third, our findings suggest that the prior opinion is also a random variable whose variance is correlated with influenceability. One can interpret the standard error of prior opinion as its standard deviation. Then, the prior opinions with larger variance tend to be more influenceable, whereas the prior opinions with lower variance are less influenceable.
On the grounds of Bayesian statistics, it is expected that weak priors are affected more by observations than strong priors, when forming posteriors, in agreement with our measurements and the prediction of self-categorization theory that the uncertainty facilitates influence. 
% future
Note, however, that while this third point is naturally explained on the grounds of Bayesian statistics, the logit model captures it only indirectly through the components of the mixture. In future works, the variance of prior opinions can be modeled with hierarchical Bayesian models and estimated with empirical Bayes methods.
% overall
Our Bayesian theory of social influence provides a base for the development of such alternative models. The theory posits that social influence is a Bayesian updating of prior opinions due to observed social feedback, possibly as a part of generic distributed Bayesian inference about the structure of reality~\cite{Krafft2016Human,Tenenbaum2011How}. 

%
%Our results suggest that the opinion formation has probabilistic nature and that the opinions are random variables with varying standard deviation, e.g., strong opinions have low standard deviation, whereas weak opinions have large standard deviation. 
%A growing body of work discusses` distributed Bayesian inference of the structure of the world\cite{Krafft2016Human,Tenenbaum2011How}. 

%% scales
%The results of our analysis are qualitatively identical and quantitatively similar independently of the psychometric scale used for surveying opinions.

%\section*{Conclusions}

% external validity
%The results of our analysis yield high external validity, because our experiments are conducted on clones of existing social media websites. We refer to this novel experimental methodology as academic A/B testing experiments. An important question is whether such experimental methods are ethical, given that participants know they joined an experiment, although they do not know that the respective social media websites are modified by researchers. Note that online platforms, such as YouTube and Facebook, routinely perform such randomized experiments (i.e., A/B testing) without informing users.

% risks to society
Opinion manipulation and misinformation are particularly pervasive~\cite{Lazer2018science,Vosoughi2018spread} and nearly effortless in online platforms, which nowadays have billions of users and become crucial for the stability of society~\cite{Richtel2013Theres,Campolo2018AI}.
% prevention
Further research is indispensable to prevent the abuse of social computing systems and to form a more robust society.
% summary of experiments
%Even online strangers influence the opinion of a considerable number of individuals.
% vulnerability
Bayesian social influence allows individuals with weak prior opinions to form more informed opinions, by the virtue of expert influence on a given topic.
However, these individuals are also vulnerable to opinion manipulation, for instance via astroturfing, i.e., paid campaigns created to influence individuals without their awareness. 
In other words, there are both good and bad effects of social influence.
% our work
Our measurements of social influence yield high external validity, because our A/B testing experiments are conducted on clones of existing social media websites. 
Our findings suggest that randomized experiments in conjunction with statistical modeling of social influence can be used to detect vulnerable users and protect them from opinion manipulation and misinformation. 
%We refer to this novel experimental methodology as academic A/B testing experiments. 
%An important question is whether such experimental methods are ethical, given that participants know they joined an experiment, although they do not know that the respective social media websites are modified by researchers. 
Note that online platforms, such as YouTube and Facebook, routinely perform such randomized experiments for other commercial purposes.
% systems
Social computing systems could be designed to emphasize the good influence and hinder the bad influence by detecting and protecting vulnerable users, and by estimating and exposing the expertise of its users within topical domains.
Although we anticipate that both domain vulnerability and expertise are related to influenceability within that domain, we point out that subsystems for measuring vulnerability and expertise shall evolve over years through an open scientific process, because of their importance to society.
Nowadays, online ratings and comments are simple and heavily affected by sampling bias, i.e., a piece of content is judged by a biased sub-sample of population and there is no way to see how other sub-populations would judge that content. Future social computing systems could characterize the people who evaluate a given piece of content, correct for the sampling bias in ratings and comments, and provide information about how experts evaluate that content. These systems would hinder opinion manipulation and the diffusion of fake news, by informing users, the vulnerable ones in particular, about the nature of ratings and comments they see.

%\showmatmethods{} % Display the Materials and Methods section

\section*{Methods}

\textbf{Comments underneath videos.}
% social feedback
%Each video comes with: i) the comments of its prior viewers and ii) the counters for views, thumbs up, and thumbs down. 
% experimental conditions
Before the experiments, from one (Experiment I) to three (Experiment II) editors label each of the comments as either positive, negative, or neutral towards the respective video. There is a significant agreement between the three labelers (Fleiss' kappa of $0.56$).
% the order of comments
The comments are always shown to participants in the reverse chronological order of their original creation date, reflecting the default setting in the respective video-sharing platforms at the time when the experiments were performed. 

\vspace{3mm}

\noindent\textbf{Experimental conditions.}
% the order of videos and experimental conditions
Each participant watches in a randomized order the same set of videos randomly assigned to the experimental or control conditions. In the case when the total number of experimental conditions is different than the number of videos, we perform a round-robin over experimental conditions to ensure a balanced assignment of conditions to videos across participants. 

\vspace{3mm}

\noindent\textbf{Psychometric scales.}
% different psychometric scales between experiments
To take robust and precise measurements, we use different psychometric scales for surveying opinions in the two experiments.
% surveying opinions and sharing willingness
The participants express their opinions about a video by declaring their agreement with the statement ``I like this video''. 
% scales
In Experiment I, the participants respond to this question on a standard 5-point Likert scale, ranging from ``Strongly disagree'' to ``Strongly agree''. 
In Experiment II, we use a more precise 200-point scale that ranges from 100\% ``Disagree'' to 100\% ``Agree''~\cite{Treiblmaier2011Benefits}.
% weighting
For the sake of simplicity, in the analysis we treat all ``Agree'' answers as positive opinion and all ``Disagree'' answers as negative opinion, independently whether they correspond to ``Weakly agree'', ``Strongly Agree'', ``5\%~Agree'', or ``75\% Agree''.
%\footnote{The results of our analysis only become more significant if we apply weighted average for computing opinions and give, for instance, twice more weight to ``Strongly agree'' than to ``Agree''.}

\vspace{3mm}

\noindent\textbf{The binomial model with decisions copying hypotheses.}
The derivation of the model follows the steps of Perez-Escudero et al.~\cite{Perez-Escudero2011Collective}, however, its new framing makes a more explicit connection to Bayesian inference, by formalizing posterior predictive probability and likelihood, and is adapted to the setting of opinion formation.
% assumptions
The derivation makes a series of simplifying assumptions, but each of them can be relaxed, as we demonstrate in the following subsections introducing other models based on the same Bayesian principles.

% main assumption
We consider an action of liking or disliking a video as a reflection of hypotheses, $\theta$, considered by the focal individual, referred to as \textit{ego} as a distinction from other individuals.
% hypotheses
Under the binomial model, we assume that ego considers the minimal number of only two hypotheses, e.g., the video is good ($\theta_+$) or bad ($\theta_-$).
%
% estimation
Ego estimates their posterior probability of each hypothesis, or posterior opinion, using their prior information about these hypotheses, $P(\theta)$, and the relevant observed social signals, $B$, with which they update their prior probability and obtain the posterior probability of hypotheses, $P(\theta|B)$, following Bayes' rule
\begin{equation}
P(\theta|B) = \frac{P(B|\theta)P(\theta)}{P(B)}.
%P(\theta|B) = \frac{P(B|\theta)P(\theta)}{P(B|\theta_-)P(\theta_-)+P(B|\theta_+)P(\theta_+)}.
\label{eq:bayes}
\end{equation}
Since only two hypotheses are considered, it is useful to write Bayes’ theorem in its posterior-odds form, that is
\begin{equation}
\frac{ P(\theta_+|B) }{ P(\theta_-|B) } = \frac{ P(B|\theta_+)P(\theta_+) }{ P(B|\theta_-)P(\theta_-) }.
\label{eq:posterior-odds}
\end{equation}
% independent opinions assumption
Next, we assume that ego estimates $P(B|\theta)$ by naively assuming that the observed opinions are independent of each other. This assumption has been shown to be a good approximation of the model including dependencies for animals~\cite{Perez-Escudero2011Collective}. 
Under this assumption $P(B|\theta) = Z \prod_{i=1}^{N}P(b_i|\theta)$, where $B$ is the set of $N$ comments read by ego and $b_i$ is the opinion expressed in the comment~$i$. $Z$ is a normalization constant ensuring $\sum_B P(B|\theta)=1$, also know as partition function, which is a combinatorial term counting the number of possible comment sequences for the set of comments $B$. 
As in the design of our experiments, we assume that each comment can be categorized as positive ($b_+$), negative ($b_-$), or neutral ($b_=$), totaling $N = n_+ + n_- + n_=$ comments. 
Then, 
\begin{equation}
P(B|\theta) = Z P(b_+|\theta)^{n_+} P(b_-|\theta)^{n_-} P(b_=|\theta)^{n_=}.
\label{eq:likelihood}
\end{equation}
We assume that neutral comments do not add any information about the correctness of hypotheses, $P(b_=|\theta_-)=P(b_=|\theta_+)$, and we will neglect them in the reminder for simplicity. Inputting the last two formulas to Equation~\ref{eq:posterior-odds} gives
\begin{equation}
\frac{ P(\theta_+|B) }{ P(\theta_-|B) } = \frac{ 
P(b_+|\theta_+)^{n_+} P(b_-|\theta_+)^{n_-}P(\theta_+) }{ 
P(b_+|\theta_-)^{n_+} P(b_-|\theta_-)^{n_-} P(\theta_-) }.
\label{eq:posterior-odds2}
\end{equation}
% log-odds
Note that $P(\theta_-|B) = 1-P(\theta_+|B)$ and the logarithm of this equation gives the log-odds
\begin{equation}
%\log \frac{ P(\theta_+|B) }{ P(\theta_-|B) } = 
\log \frac{ P(\theta_+|B) }{ 1-P(\theta_+|B) } =
n_+ s_+ + n_- s_- + a,
\label{eq:log-odds}
\end{equation}
where $s_+ = \log \frac{P(b_+|\theta_+)}{P(b_+|\theta_-)}$, $s_- = \log \frac{1-P(b_+|\theta_+)}{1-P(b_+|\theta_-)}$, and $a=\log \frac{P(\theta_+)}{P(\theta_-)}$.
The parameter $a$ captures the relative prior probability of the two hypotheses, i.e., the relative prior opinion of ego,\footnote{In our terminology, prior and posterior opinions are synonyms of prior and posterior distributions over hypotheses, whereas expressed opinions are samples from these distributions.} whereas $s$ determines how much is the prior opinion affected by the observed social signals.
This formula for log-odds is further simplified, if we assume a symmetric influence of positive and negative comments, i.e., if $s_+ = -s_- = s$, then a positive comment negates a negative comment. 
We recognize that the log-odds in Equation~\ref{eq:log-odds} are a linear function of the observed $n_+$ and $n_-$, so its parameters $s$, and $a$ can be estimated with a logistic regression model
\begin{equation}
P(\theta_+|B) = \sigma( s \Delta n + a ),
\label{eq:posterior_logistic}
\end{equation}
where $n=n_+-n_-$ and $\sigma$ is a logistic function, but we still need to relate this posterior probability of hypotheses to the opinions expressed by ego. 

%\textbf{The decision rule and its discussion.}
%In the next section we consider different decisions rules that ego can apply to express their opinion based on the estimated posterior over hypotheses, $P(\theta|B)$.
% decision rule - prior matching
%The remaining import question is how decisions, to express an opinion, are made, i.e., what decision rule is used, given the posterior probability of hypotheses. 
So far we have considered the perceptual stage of decision-making, in which ego estimates which of the hypotheses is correct. 
Whether the video is liked or not is decided by a decision rule. 
Evidence for animals and humans suggests that individuals use a decision rule called probability matching \cite{Behrend1961Probability, Kirk1965Probability, Wozny2010Probability, Perez-Escudero2011Collective, Arganda2012common}. According to this rule, the ego expresses an opinion, $b$, by directly drawing the corresponding hypothesis from the posterior distribution of hypotheses, i.e.,
\begin{equation}
P(b=b_+|B) = P(\theta_+|B).
%= P_+(\Delta n,s,a)
\label{eq:prob_match}
\end{equation}
% In other words,
% \begin{equation}
% P(b|B) = \delta_{bb_+} P(\theta_+|B) + \delta_{bb_-} P(\theta_-|B),
% %= P_+(\Delta n,s,a)
% \end{equation}
% where $\delta_{ij}$ is Kronecker delta, i.e., $\delta _{ij}=0$ if $i\neq j$.
%
This decision rule is equivalent to the typical rule that future observations are draws from the posterior predictive distribution, that is the likelihood averaged over the posterior
\begin{align}
P(b=b_+|B) &= 
\sum_{ \theta \in \{\theta_+, \theta_-\} } P(b=b_+|\theta) P(\theta|B),
\label{eq:posterior_pred}
\end{align}
if only there is one-to-one mapping between $b$ and $\theta$ and samples of $b$ from the likelihood are copies of the draws from the corresponding posterior distribution of $\theta$, i.e., $P(b|\theta)=1$ if $b=b_+$ and $\theta=\theta_+$ or $b=b_-$ and $\theta=\theta_-$; otherwise $P(b|\theta)=0$.
Thus, there is a direct mapping between hypotheses and expressed opinions; the difference between the two is that expressed opinions are drawn at random at the moment of an observation, whereas hypotheses are latent opinions that are not observed directly until they are copied and expressed.
%$P(b=b_+|\theta)=\delta_{\theta_+\theta}$, where $\delta_{ij}$ is the Kronecker delta taking 1 if $i=j$ or 0 otherwise.
Finally, note that when ego is deciding what opinion to express (Equation~\ref{eq:prob_match}), the likelihood function copies a draw from the posterior of ego; however, when the individuals whose opinions are observed by the ego are making a decision (Equation~\ref{eq:likelihood}), then the likelihood function copies a draw from their own distributions over hypotheses, which ego aims to estimate with the parameter $s$.

\vspace{3mm}

\noindent\textbf{Model fitting.}
% inference
The parameters of a mixture of logit models can be inferred with the expectation-maximization algorithm, but this approach gets stuck in local optima~\cite{Sedghi2016Provable}.  
Thus, we use a different, approximate, method for inferring the parameters of each model.
Namely, we treat each individual as indistinguishable and estimate the probability of positive opinion as $P_{+}(\Delta n)= \sum_{k=1}^{K} p_k P_{+}(\Delta n, s_k, a_k)$, where $p_k$ is the fraction of individuals in the sub-population~$k$ and $K$ is the total number of sub-populations. This probability does not depend on any particular individual, but instead it averages the probability of positive opinion over all individuals. 
%The values of model parameters are inferred by maximizing likelihood (see The Model of Sub-populations in SI Appendix).

% inference
The parameters of a mixture of logit models can be inferred with the expectation-maximization algorithm, but this approach gets trapped in local optima~\cite{Sedghi2016Provable}. Thus, here we use a mean-field approximation of that model to find optimal values of parameters. Namely, we assume that individuals are indistinguishable. In such case, the probability that an unidentifiable individual from the whole population has a positive opinion about the video is
\begin{equation}
P_{+}(\Delta n)= \sum_{k=1}^{K} p_k \frac{1}{1+\exp(-s_k \Delta n - a_k)},
\label{eq:model_macro}
\end{equation}
where $p_k$ is the portion of individuals in sub-population $k$. 
%This equation is a mean-field approximation of the probability of positive opinion for an individual if we do not know to what sub-population this individual belongs to.
% joint probability
The joint probability of observing opinions $\matr y$, given that the individuals were exposed to $\matr \Delta n$ comments is
\begin{equation}
\begin{split}
L  
&= \prod_{u=1}^U \prod_{v=1}^V
\left( P_{+}\left(\Delta n_{uv}\right) \right)^{y_{uv}} 
\left( 1-P_{+}\left(\Delta n_u\right) \right)^{1-y_{uv}}
%&= \prod_{u=1}^U \prod_{v=1}^V
%\left( \sum_{k=1}^{K} \frac{p_{kv}}{1+\exp(-s_{kv} \Delta n - a_{kv})} \right)^{y_{uv}} 
%\left( 1-\sum_{k=1}^{k} \frac{p_{kv}}{1+\exp(-s_{kv} \Delta n - a_{kv})} \right)^{1-y_{uv}} \\
%&\equiv P(\matr {\Delta n}, \matr y | K, \underset{K \times V}{\matr{p}}, \underset{1 \times V}{\matr{a_1}}, \underset{1 \times V}{\matr{s_1}}, \dots, \underset{1 \times V}{\matr{a_K}}, \underset{1 \times V}{\matr{s_K}}  )
\label{eq:p-complete}
\end{split}
\end{equation}
where $U$ is the total number of individuals and $V$ is the total number of videos, and ${\sum_k p_{kv} = 1}$ for each video $v$.
To fit the parameters of this model, we maximize the log-likelihood $\log(L)$. We present detailed results of this fitting in the following subsections.

\vspace{3mm}

\noindent\textbf{Standard errors of parameters.}
% standard errors of params
The standard error of each estimated parameter are obtained using three different methods. The first two methods correspond to random re-sampling of results among individuals: either via bootstrapping or jackknife approach. In the bootstrapping approach, we sample the results of experiment with replacements to obtain the same number of samples, that is individuals, as in the original experiment. Then, we fit the parameters of the model using such re-sampled data. We repeat this procedure $1000$ times and compute the standard error of the estimated parameters. The jackknife approach follows the same procedure, except that instead of sampling, we randomly drop one sample from the set of original results of the experiment. The third method of computing standard error is based on the analysis of the log-likelihood. We note that the covariance matrix of the estimated parameters $\hat{\theta}$ is an inverse of the observed Fisher information (negated Hessian of log-likelihood):
\begin{equation}
\pmb{\Sigma}( \hat{\theta} ) =  
\left[ \pmb{\mathcal{I}} ( \hat{\theta} ) \right]^{-1} = 
\left[ -\frac{ \partial^{2} \left( \log \left( P \left( \matr {\Delta n}, \matr y | \theta \right) \right) \right)  }
{\partial\theta_{i}\partial\theta_{j}}
\bigg|_{ \theta = \hat{\theta} } \right]^{-1}.
\end{equation}
Thus, the standard error of an estimated parameter $\hat{\theta}_k$ is the square root of the corresponding diagonal element of $\pmb{\Sigma}( \hat{\theta} )$. We compute the standard error for each parameter using this method and report it in the main text.

%We observe a couple of patterns in the values of parameters. 
%First, the influenceable sub-population is smaller than the non-influenceable one, i.e., the parameter $p_{2v} \in (0.05,0.4)$. This result suggests that over half of participants was not influenced by the social feedback, whereas a considerable fraction was influenced. Second, 
Some of the parameters differ considerably between videos, especially the prior opinions $a_k$ about the videos. Interestingly, the prior opinion of the non-influenceable sub-population about a given video takes similar values in various top models, whereas the prior opinion of influenceable sub-population varies largely across top models. Also, the standard error of the prior opinions is many times larger for the influenceable than non-influenceable sub-population.
%One can hypothesize that social influence depends on the topic of the video and the topical expertise of the audience that watches it.
This difference may arise due to the fact that the non-influenceable sub-population is larger than influenceable sub-population. If we interpret the prior opinion $a_k$ of sub-population $k$ as a mean over prior opinions $a_{ku}$ of individuals belonging to this sub-population, then the standard error of this mean is $\sigma( a_k ) = \sigma( a_{ku} ) / \sqrt{U_k} $, where $U_k$ is the number of individuals belonging to that sub-population. Thus, to compare the standard errors of this parameter for the two sub-populations, we shall compute the ratio 
\begin{equation}
\frac{ \sigma( a_{2u} ) }{ \sigma( a_{1u} ) } = 
\frac{ \sigma( a_{2} ) \sqrt{U_2} }{ \sigma( a_{1} ) \sqrt{U_1} }  = 
\frac{ \sigma( a_{2} ) \sqrt{p_2} }{ \sigma( a_{1} ) \sqrt{p_1} },
\end{equation}
where $p_k$ is the fraction of individuals belonging to the sub-population $k$. This ratio compares the intrinsic standard deviations of prior opinion of individuals belonging to two different sub-populations, taking into account that they differ in size. 
%With this ratio, we show that individuals in influenceable sub-population have much noisier prior opinions about each video than individuals in non-influenceable sub-populations (Figure~\ref{fig:bestparams}).

\vspace{3mm}

\noindent\textbf{Confidence intervals of the running probability.}
% running probability
For the purpose of the presentation, we compute running probability of positive opinion about a video. 
Namely, given a set of answers $y$ from different users for specific  $\Delta n$, we compute running probability of positive opinion with a sliding window of $n$ data points. 
%This \textit{shuffled} running probability is the mean of $m$ normal running probabilities for random permutations of the set \textbf{$y$} for each value of $\Delta n$. 
To compute this running probability, the set of answers $y$ is ordered in the increasing order of $\Delta n$. Then, for every \textit{i}-th window of $n$ experimental points we compute $\overline{\Delta n_i} = \frac{1}{n}\sum_{u=1}^{n} \Delta n_{u+i-1}$ and $P_{+, i,j} = \frac{1}{n} \sum_{u=1}^{n} y_{u+i-1} $.
In our dataset, for a given value of $\Delta n$ usually there are several answers from different participants for which $\Delta n$ is the same. 
Note that the answers for a given $\Delta n$ do not have a natural order. 
%The arbitrary order given to compute a normal running probability would give one of the possible values of $P_{+ i}$. This value depends on the arbitrary order assigned and it does not reflect all the possible orders of the data. 
To avoid any artifacts in the computation of the running probability due to the lack of order in the answers for a given $\Delta n$, we randomly permute the answers for that $\Delta n$ and compute $P_{+, i, j}$, where $j$ stands for that permutation. We repeat this process $m$ times to obtain the final running probability of positive opinion as an average over $m$ permutations, i.e., $\overline{P_{+,i}} = \frac{1}{m} \sum_{j=1}^{m} P_{+,i,j}$.

% ci
To show how well the model predicts the experimental probability, we compute the $99\%$ confidence interval of the model for the running probability. To this end, we calculate the running probabilities based on the artificial data simulated with the fitted model for the real finite set of $\Delta n$. We repeat this procedure $1000$ times to obtain $99\%$ confidence intervals of the model for each $\overline{\Delta n_i}$. We present these confidence intervals in Figure 1 of the main text and all other figures of running probability.

\acknow{We thank Luis Fernandez Lafuerza for insightful discussions and feedback, Karin Chellew for her suggestions on the design of the experiments, Maria Cano-Colino for her help with comment labeling, and Manuel Gomez-Rodriguez for his comments on model learning. We acknowledge funding from the Volkswagen Foundation (P.A.G.), the Alexander von Humboldt Foundation (F.B.), the Funda\c c\~ao de Amparo  \`a Pesquisa de Minas Gerais (F.B.), the Funda\c c\~ao para a Ci\^encia e Tecnologia PTDC/NEU-SCC/0948/2014 (G.G.d.P.), and the Champalimaud Foundation (G.G.d.P.).}

\showacknow{} % Display the acknowledgments section

% Bibliography
\bibliography{jabref,manually_added}

\renewcommand{\thefigure}{S\arabic{figure}}
\renewcommand{\thetable}{S\arabic{table}}

\clearpage
\part*{Supporting Information Appendix}
\tableofcontents
%\clearpage

\section{Experiment I}
\label{sec:preliminary}

% ===== introduction to our experiment and the paper
We first conducted Experiment I, in preparation to Experiment II. 
To each participant, we showed videos with corresponding social feedback from two video-sharing websites, i.e., YouTube and Vimeo. We asked the participants what is their opinion about each of the videos.
In Experiment I, we used a 5-point Likert scale to measure the opinion and we introduced \textit{partially} manipulated conditions in which we altered either only comments or only the counters for thumbs and views.
Here, we present a detailed description of this experiment and its results.

\subsection{Description}

\begin{figure*}[htbp]
\centering
\includegraphics[width=0.7\textwidth]{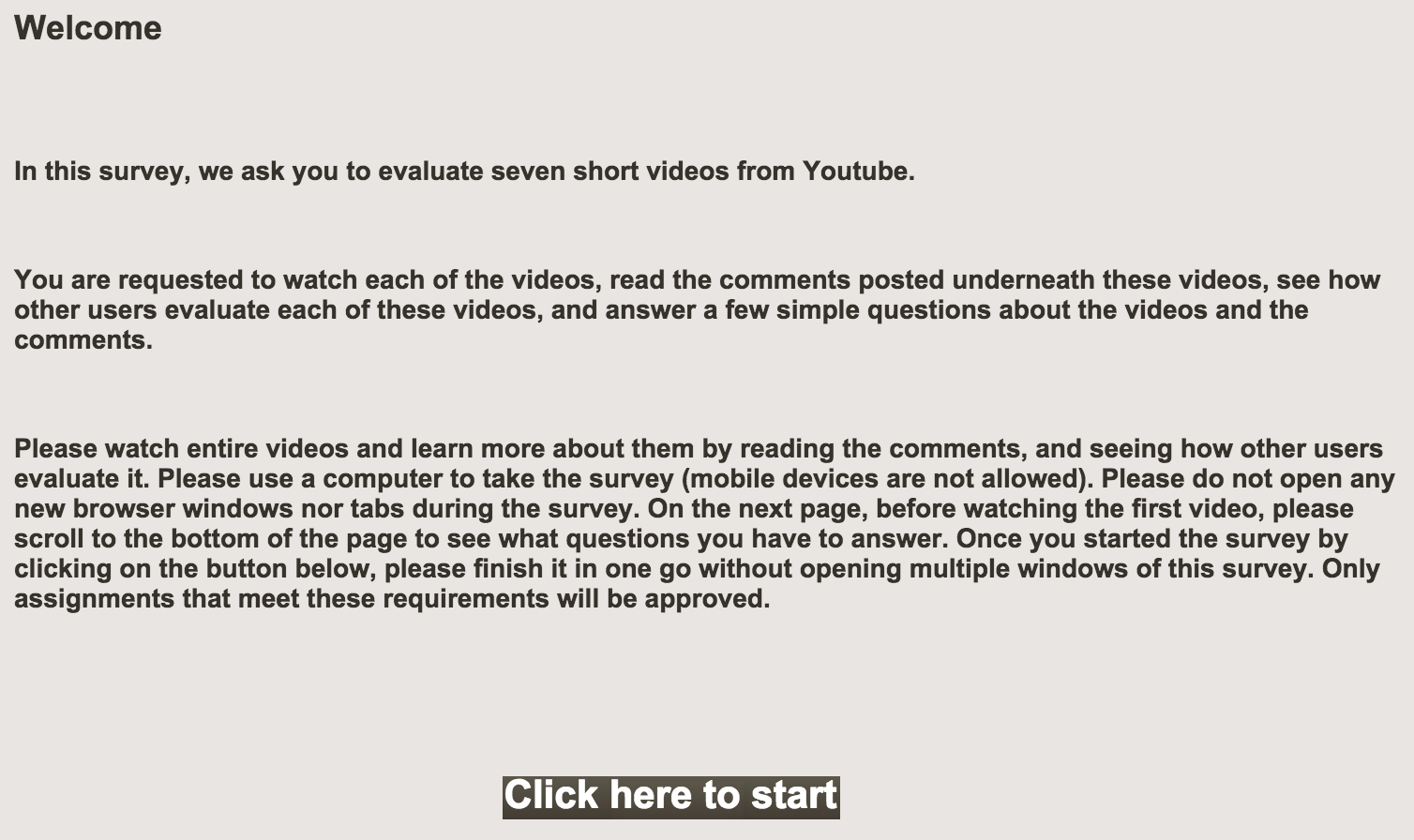}
\caption{Instructions given before starting the survey.}
\label{fig:screenshot_instructions} 
\end{figure*}

\begin{figure*}[hbtp]
\centering
\includegraphics[width=0.58\textwidth]{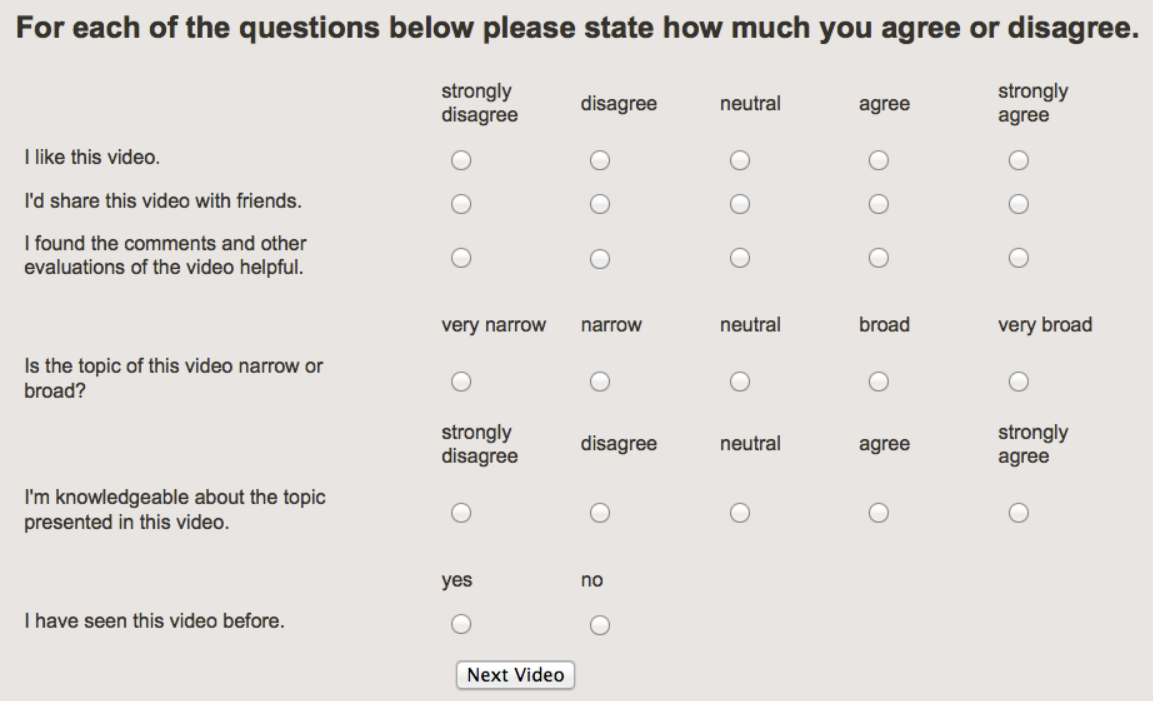}
\caption{The survey performed immediately after the participant watched the video.}
\label{fig:survey}
\end{figure*}

Before starting the survey the participants are given the instructions (nearly identical for both experiments, Figure~\ref{fig:screenshot_instructions}).  
The subjects are asked to click on the image to watch the video, see the feedback of other people, and to evaluate the video. For each video, we ask a set of questions about the video (Figure~\ref{fig:survey}). The participant can see the questions before watching the video, but cannot answer them without playing the video first, and all questions must be answered before advancing to the next video. 
% ===== the social feedback and the intro to the experimental conditions
%The participants of the experiment are asked first to watch a video and then to read the feedback of other users about the video. 
The pages with the videos look exactly like in the original systems, i.e., in Youtube and Vimeo, except the social feedback to the videos is modified to various extent under different experimental conditions. 
The comments are shown under videos in the reverse chronological order of their original creation date, reflecting the default setting in the respective video-sharing platforms at the time when the experiments were performed. 
%In Vimeo there is no thumbs down, so only the comments and the numbers of thumbs up and views are modified. 

% ===== the questions and the quality of responses checks
For each video we conduct a survey (Figure~\ref{fig:survey}). We ask the participant on a $5$-point Likert scale if she agrees with the statement evaluating her opinion about the video (``I like this video'') and her willingness to share the video (``I'd share this video with friends'').
%For the purpose of presentation, we translate the ordinal $5$-point responses of participants, ranging from ``strongly disagree'' to ``strongly agree'', to integer numbers ranging from $-2$ to $2$, respectively.
Additionally, we ask if the person saw this video before. 
% ===== correlation of opinion and rating
%Note that the opinion can be interpreted as a 5-star rating. Pearson correlation between the average opinion about video in the control condition and the ratio of thumbs up to the total number of thumbs is $0.81$ for YouTube videos, although this result is not significant due to the small number of videos ($p=0.19$). \pms{Compute that for full experiment}
%
In the analysis, we discard the responses which were influenced by the past exposures to the given video. %Furthermore, we discard all video session responses for which the time spent during the session was lower than the length of the video. Furthermore, we discard all surveys for which the total time spent on the experiment was lower than the total length of all videos. 
%After performing all the mentioned discards we retain $55\%$ of all initial responses, namely, over $260$ responses per each condition per each of two systems.
%(the exact numbers of responses are given in Table~\ref{tab:counts}). 
%The details of the experiment setup are described in Appendix A (the representative parts of the experiment are shown in Figure~\ref{fig:survey}).

\subsection{Videos}

\begin{table*}[tbh!]
\centering
\begin{tabular}{c|c|ccc|cc|c|c}
\toprule
& & \multicolumn{3}{c|}{Comments} & \multicolumn{2}{c|}{Thumbs} & \\
Platform&             Video &  Pos. &  Neut. &  Neg. &   Up & Down &    Views &                                 Link to the original \\
\midrule
 &           HTC add &        17 &       30 &        47 &  110 &   31 &  147,273 &  \href{http://youtu.be/dwGGdM3Nj08}{dwGGdM3Nj08} \\
YouTube &       Lamborghini &        14 &        1 &         5 &   41 &   28 &   76,675 &  \href{http://youtu.be/Pc7XHHCjtJI}{Pc7XHHCjtJI} \\
 &    A girl singing &        37 &       15 &        37 &  663 &  525 &  120,938 &  \href{http://youtu.be/5JKJhY15NNA}{5JKJhY15NNA} \\
 & Supermarket joke &         5 &        3 &        15 &   60 &   45 &   25,892 &  \href{http://youtu.be/VqGaHxC3Zbg}{VqGaHxC3Zbg} \\
\midrule
 &  Google joke &         2 &        7 &        10 &   50 &    - &        - &    \href{https://vimeo.com/9261909}{9261909} \\
Vimeo &        Stride add &         8 &        8 &         3 &  221 &    - &        - &   \href{https://vimeo.com/23061242}{23061242} \\
 &   Ipad skateboard &        33 &       14 &        13 &  842 &    - &        - &   \href{https://vimeo.com/11480457}{11480457} \\
 &     Curing cancer &        19 &       10 &         8 &  695 &    - &        - &   \href{https://vimeo.com/54668275}{54668275} \\
\bottomrule
\end{tabular}

\caption{Basic statistics of the original (non-manipulated) videos: the number of comments of various types, the number of thumbs up and thumbs down, and the number of views at the time of their collection (January 2014).}
\label{tab:pre-videostats}
\end{table*}

% ===== the videos
For the experiments, we picked four videos from YouTube and four videos from Vimeo. To avoid videos that subjects have seen before, we chose videos that have a public appeal but are not very popular, namely have from $20,000$ to $1$ million views, more than 15 comments (including some positive and negative ones), and over 15 thumbs. The links to the original videos and the numbers of comments, thumbs, and views for each of the videos are listed in Table~\ref{tab:pre-videostats}. The average duration of videos is $104$ seconds.
% ===== highl-level view of the experiment
During the experiment each participant is shown all eight videos in a random order, each of them under a different experimental condition, chosen in a round-robin fashion from: the control condition, two strongly modified conditions, two mild conditions, two weak conditions, and four partial manipulations. We describe the experimental conditions below.

\subsection{Comments}
Each of the comments was labeled by an author as either positive, neutral, negative, or unreadable (see Table~\ref{tab:pre-videostats} for a summary). 
Positive comments are the comments that describe the video in a positive way, while negative comments are negative toward some aspects of the video. Neutral comments are mostly off-topic or do not contain any evaluations of the content of the video. 
The comments that are unreadable or are written in a language different from English are filtered out. 

\subsection{Data Processing}

Before analyzing the data, we clean and filter it. Namely, we discard all answers that are incomplete due to technical reasons or individual mistakes, as well as double answers from the same participant (in total, we discard less than $6\%$ of all participants). 
%Then, we apply the following filters. 
To estimate the quality of answers, we measure how much time it takes for each participant to evaluate each video. We exploit this information, to invalidate the video evaluations that took a subject less time than half of video duration. Furthermore, we also invalidate the evaluations of videos that have been seen by the participant externally before the experiment, according to self-reports of participants for each video.
Then, we discard the participants with answers invalidated for more than one videos. In total, we discard less than $10\%$ of all participants.

\subsection{Main Experimental Conditions}

The main experimental conditions include the control condition, two strongly modified conditions, two mild conditions, and two weak conditions.
%Our control condition consists of the original video with the original social feedback. In a positively (negatively) manipulated condition we hide a part or all of the negative (positive) comments, increase (decrease) the number of thumbs up and the number of views. In short, in the manipulated conditions we show the same page with social feedback changed accordingly.
% intro to extreme manipulations
In \textit{strongly modified} positive (negative) condition, we hide all negative (positive) comments, we show all neutral and positive (negative) comments, and we increase (decrease) the numbers of thumbs up and views by the factor of $10$. The mildly and weakly modified conditions are implemented as \textit{m-factor manipulations}, which fine-tune the extent of modifications with respect to the control condition. In the m-factor positive (negative) manipulation, we hide all negative (positive) comments except randomly chosen $1/m$ comments, and multiply (divide) the numbers of thumbs up and views by the factor $m$. 
The mild conditions are $6$-factor manipulations, whereas the weak conditions are $3$-factor manipulations.
%Now, to study how opinion changes with the amount of past positive and negative feedback, in addition to the extreme manipulations, we explore four other experimental conditions, namely 3-factor and 6-factor positive and negative manipulations.
Note that the aforementioned strongly modified conditions correspond to $10$-factor manipulations, except \textit{all} comments of certain type are hidden instead of their majority. 

%\begin{table*}[hbtp]
%\centering
%\input{include/youtube_split-timefix1-correct-a_pvalues-grad_r1.tex}
%\caption{Statistical significance, according to Mann-Whitney U test, against the alternative hypothesis that under specified condition (column) the respondents liked the YouTube videos less than under another condition (row). For corresponding results for Vimeo see Table~\ref{tab:p:vgrad1}.}
%\label{tab:p:ytgrad1}
%\end{table*}

\subsection{Partial Experimental Conditions}

% ===== intro to manipulations - part
We introduce \textit{partial} experimental conditions to find which type of social feedback has larger influence on opinion change. Partial experimental conditions are the same as the strong conditions, except they alter either only comments or only the counters of thumbs and views, while keeping the other unchanged with respect to the control condition. 
Namely, in the partial positive (negative) condition manipulating comments, we hide all negative (positive) comments, while keeping the numbers of thumbs and views unchanged. %The comments are modified in partial conditions in the same way as in the strongly modified conditions. 
%The only difference between the extremely manipulated condition and the condition with manipulated comments is the manipulation of thumbs up and views. 
In the partial positive (negative) condition manipulating counters, we multiply (divide) the number of thumbs up and views by $10$, while keeping the comments unchanged. These experimental conditions allow us to measure whether the comments or the counters impact opinions more.

\subsection{Results}

%\pms{These numbers are for YouTube only, merge them with the results for Vimeo.}

\begin{figure}[t]
\begin{picture}(290,135)
\centering
\put(0,0){
\includegraphics[width=0.24\textwidth]{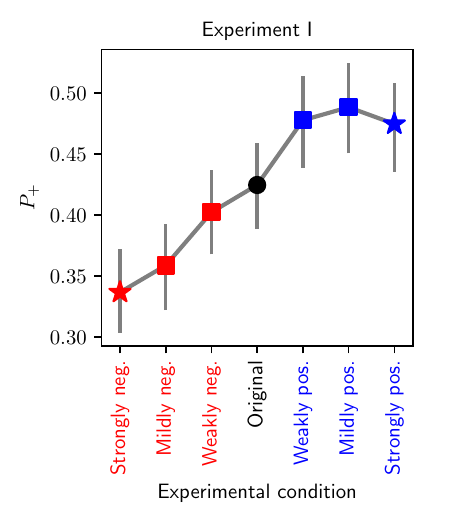}
\includegraphics[width=0.24\textwidth]{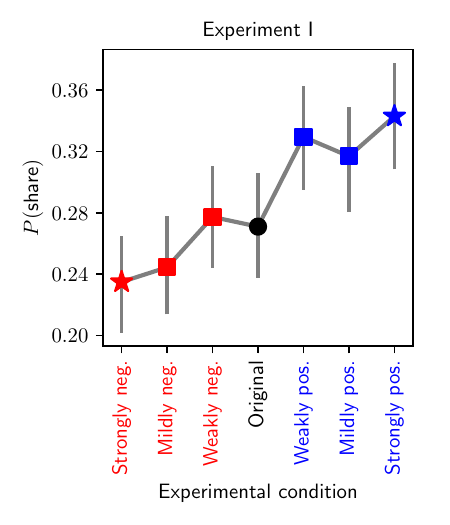}
}
\put(33,120){\small A}\put(160,120){\small B}
\end{picture}
\caption{The probability of positive opinion (left) and the sharing willingness (right) for the control condition (black circles) and the experimental conditions of various strength. The color of markers encodes positive (blue) and negative (red) conditions.}
\label{fig:exp1grad}
\end{figure}

% ===== the results - grad
We compute the probability of positive opinion (Figure~\ref{fig:exp1grad}A) and the willingness to share the video (Figure~\ref{fig:exp1grad}B) under each of the seven main conditions. 
We find that there is a statistically significant difference between the opinions about videos shown under positive and negative manipulations (Mann-Whitney U test, $p<10^{-9}$), as well as in comparison with the control condition ($p=0.007$ and $p<10^{-3}$, respectively). Similarly, we find significant difference in sharing likelihood between the positive and negative extreme manipulations ($p < 10^{-5}$), and in comparison with the control condition ($p = 0.001$ and $p=0.047$). Therefore, we find the evidence of opinion change, as well as the change in sharing willingness, in the case of both positive and negative manipulations.
In fact, the differences are significant for most pairs of experimental conditions ($p<0.05$).

%
%\begin{figure}[t]
%\centering
%\subfigure{\includegraphics[width=0.24\textwidth]{figs2/survey2-a_treat-correct_resp-avg-opinion-part_bothsolo}}
%\subfigure{\includegraphics[width=0.24\textwidth]{figs2/survey2-a_treat-correct_resp-avg-share-part_bothsolo}}
%\caption{The probability of positive opinion (left) and the sharing willingness (right) for the control condition (black circles) and the experimental conditions in which either comments or counters or both are modified. The $95\%$ confidence intervals are from BCA bootstrap.}
%\label{fig:part}
%\end{figure}

\subsection{Demographics of Participants}

\begin{figure}[btp]
\begin{picture}(290,100)
\centering
\put(0,0){
\subfigure{\includegraphics[width=0.24\textwidth]{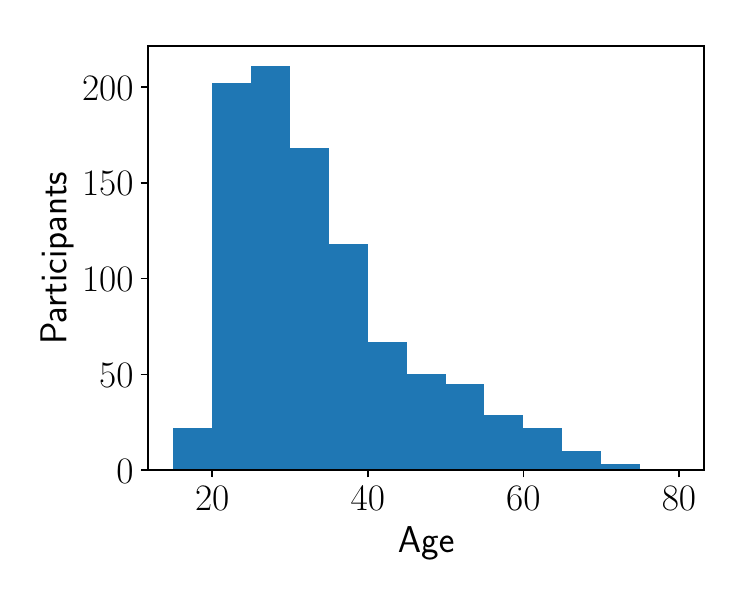}}
\subfigure{\includegraphics[width=0.24\textwidth]{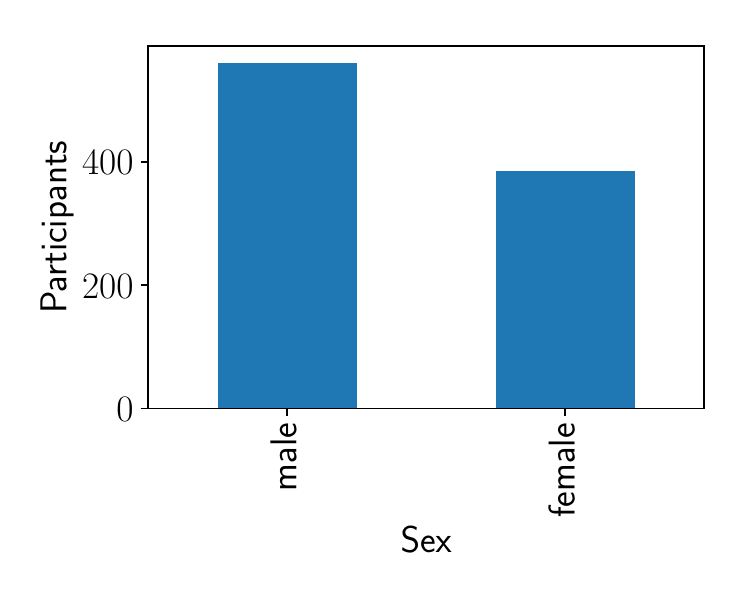}}
}
\put(110,82){\small A}\put(234,82){\small B}
\end{picture}
\caption{Demographic information about the participants of the experiment.}
\label{fig:exp1demographics}
\end{figure}

The participants of the experiment were recruited via Amazon Mechanical Turk. They performed the survey anonymously, voluntarily, and were compensated. At the end of the experiment, they answered a couple of questions about their demographics. The participants of this experiment are more likely to be male than female and are predominantly young adults (see Figure~\ref{fig:exp1demographics}).

\section{Experiment II}
\label{sec:experiment}

\begin{figure}[htbp]
\centering
\includegraphics[width=0.5\textwidth]{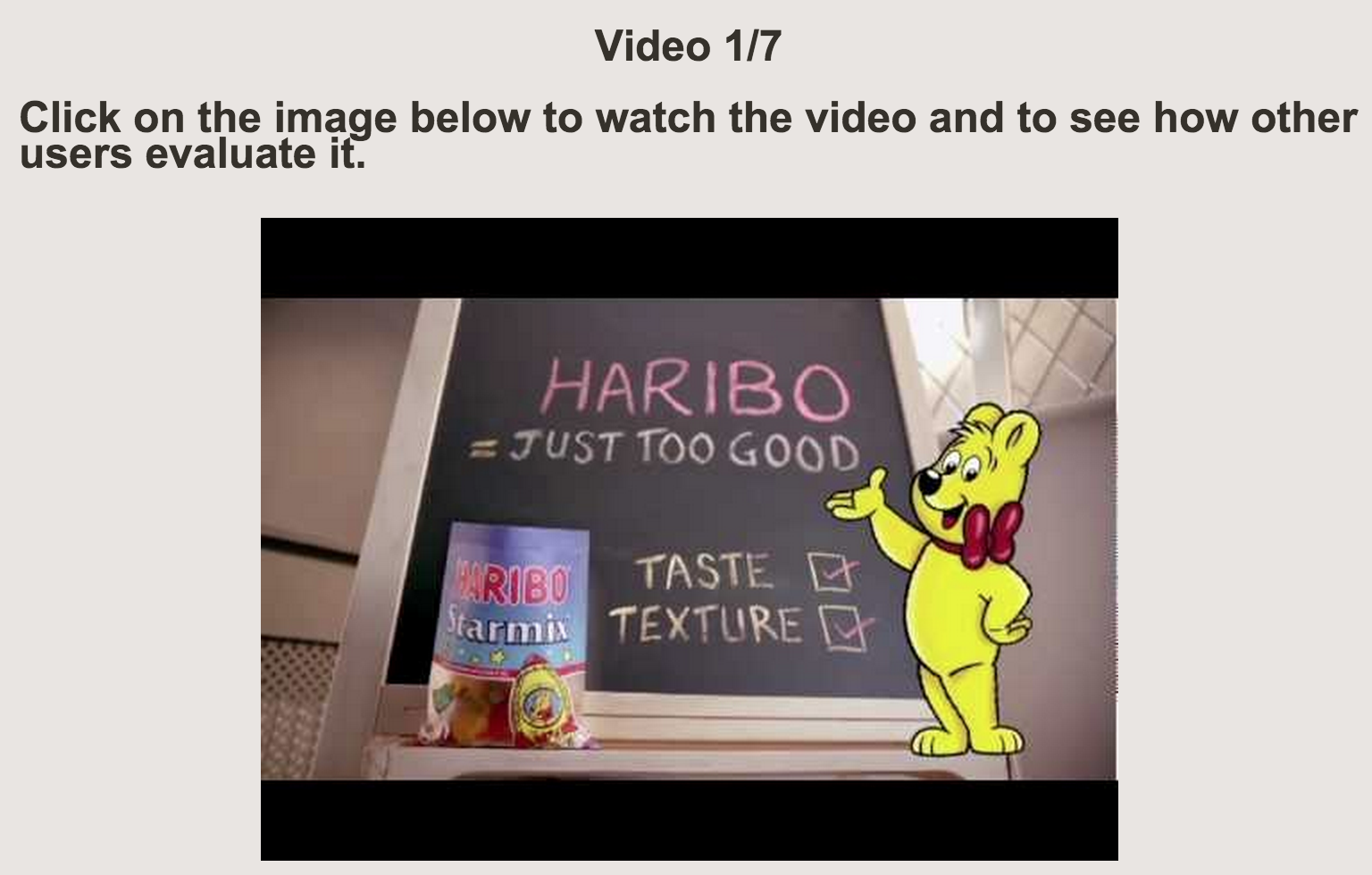}
\caption{Participants access a video on YouTube or Vimeo by clicking on its thumbnail.}
\label{fig:screenshot_survey1}
\end{figure}

% ===== introduction to our experiment and the paper
%Overall, we conduct a large-scale experiment involving over $1,400$ participants. 

In this experiment, we track which exact comments are shown on the screens of the participant of the experiment and we use a sophisticated 200-point scale for measuring opinion. This experiment has two parts. Each of the parts is conducted with a different set of 7 videos on a different set of 700 participants. We split this experiment in two parts to avoid overloading the participants with evaluations of too many videos.
In each part of this experiment, a participant is shown $7$ videos in a random order in $7$ different main experimental conditions assigned randomly to the videos.

%For the purpose of this study, two surveys have been performed on Amazon Mechanical Turk (AMT). Each survey consists of seven videos from YouTube. Each of the two surveys was finished by 700 participants. 
%To each participant, we show videos with their original social feedback from the video-sharing website where they were posted (i.e., YouTube). We ask the participants what is their opinion about each of the videos with five questions. The original social feedback shown on the page of each of the videos is manipulated to a various extent under different experimental conditions, while the content of the videos remains unchanged. For each of the conditions, we measure the change of opinion of the viewers caused by the exposure to the altered feedback. We present the results and the details of the methods and materials in the following subsections.

\subsection{Description}

\begin{figure*}[htbp]
\centering
\includegraphics[width=0.7\textwidth]{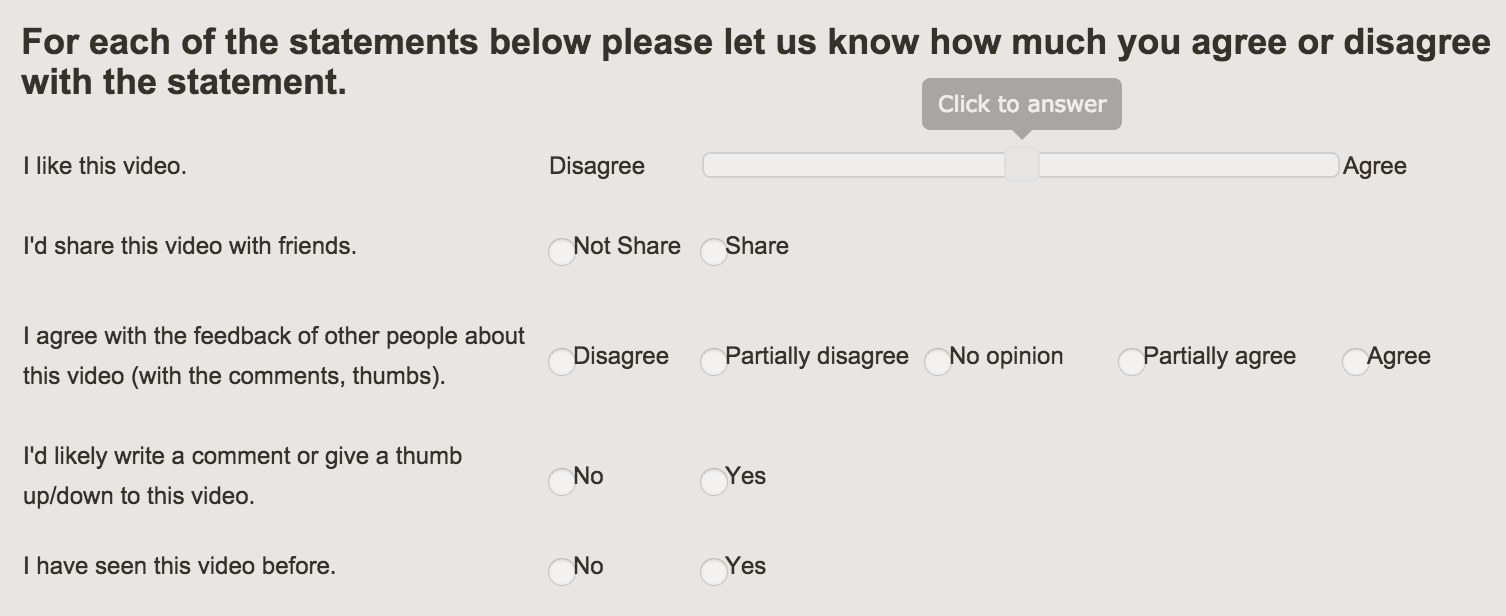}
\caption{Questions asked for each video. The opinion is measured with the statement ``I like this video''. } 
\label{fig:screenshot_survey2}
\end{figure*}

\begin{figure}[tbp]
\centering
\includegraphics[width=0.35\textwidth]{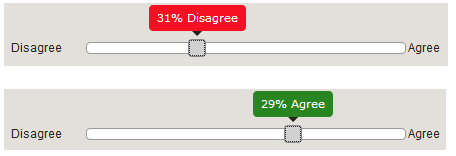}
\caption{The psychometric scale for measuring opinion. The opinion is measured with the statement ``I like this video''. Subjects answer this statement with a slider ranging from $100\%$ ``Disagree'' to $100\%$ ``Agree''. } 
\label{fig:screenshot_psychometric}
\end{figure}

\begin{figure*}[htbp]
\centering
\includegraphics[width=0.65\textwidth]{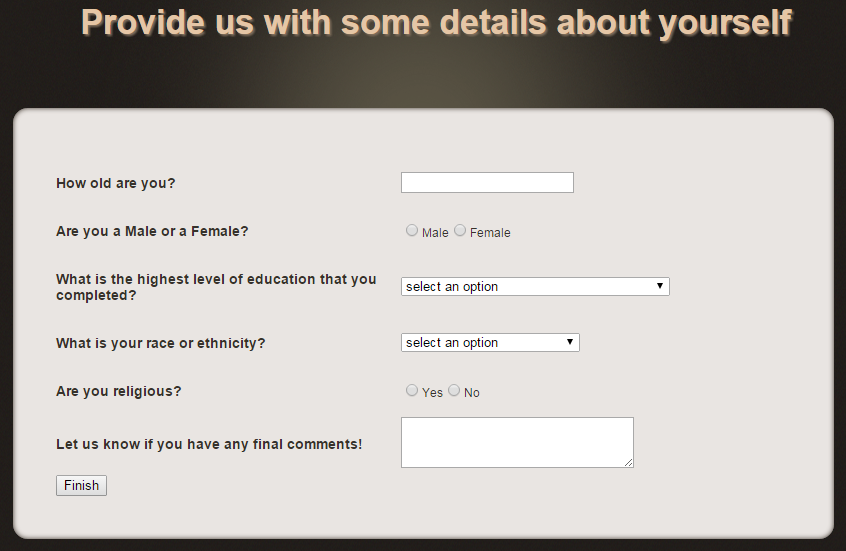}
\caption{Demographic questions. After completing the survey for the seven videos the participant was asked to answer the following questions about personal demographic information.}
\label{fig:demographic_survey}
\end{figure*}

The participants are given the same instructions as in Experiment I. 
They are instructed to watch videos, to look at the feedback from other people about these videos, and to evaluate them.
The subjects enter a webpage that looks like YouTube by clicking on a thumbnail of the video (Figure~\ref{fig:screenshot_survey1}). 
%For each video in the survey there were five questions about the video (Figure~\ref{fig:screenshot_survey2}).
For each video shown in the experiment, we ask the participant to evaluate whether she agrees or disagrees with the following statements (Figure~\ref{fig:screenshot_survey2}). To measure the opinion, we ask about the agreement with the statement ``I like this video''. The participant answer this statement using a slider bar in a scale from $100\%$ to $100\%$(Figure~\ref{fig:screenshot_psychometric}). We did not give the option of answering $0\%$ to avoid neutral answers. To measure the sharing willingness, we use the statement: ``I'd share this video with friends'' 
%and ``I'd like to write a comment or give a thumb up/down to this video''. These statements can be answered with ``Not Share/Share'' and ``Yes/No'', respectively. Note that the answer to these statements does not necessarily correlates with the opinion of the participant. For example, the participant could write either positive or negative comment. 
This statement can be answered with ``Not Share/Share''.
To test whether the participant agrees with other people's feedback, we use the statement ``I agree with the feedback of other people about this video (with the comments, thumbs)''. 
%Subjects' comments about the preliminary experiment suggest to use a finer scale for the agreeableness with this statement.}
%\paco{I'd substitute this paragraph with the previous one} For each video shown in the experiment, we ask the participant with a slider if she agrees with the statement evaluating her favorableness toward the video ("I like this video") and her inclination toward sharing the video ("I'd share this video with friends"). While the first question evaluates the opinion about the video, the second question estimates the probability of an action related to that opinion.
Additionally, we ask if the person have seen this video before, to account for the possibility of prior influence outside of the experiment. 
After having answered these questions for each of the videos, the participant is asked a short demographic survey (Figure~\ref{fig:demographic_survey}).

%\begin{figure}[htbp]
%\centering
%\includegraphics[width=0.5\textwidth]{figs/screenshot_survey1.png}
%\caption{\textbf{Access to the video.}}
%\label{fig:screenshot_survey1}
%\end{figure}

%\begin{figure*}[hbtp]
%\centering
%\includegraphics[width=0.60\textwidth]{figs/surveyscreen1b}
%\caption{The pages of the experiment showing the video to watch, which looks exactly like the original YouTube page of that video, with all the comments to this video posted underneath of it.}
%\label{fig:ytvideopage}
%\end{figure*}

\subsection{Videos}

For the experiments, we picked fourteen videos from YouTube of diverse quality, as judged by the fraction of thumbs up among thumbs (see Table~\ref{tab:videostats}). Note that in the main text, the fraction of thumbs up is used as an estimate of external opinion.
We chose videos that have a public appeal but are not very popular, namely have from $100,000$ to $2$ million views, more than 50 comments, and over 100 thumbs (at the moment of data collection). Since we do not create any artificial comments and use only the original comments to manipulate the social feedback, we picked only videos that have both positive and negative comments. The links to the original videos and the numbers of each of the comment types, thumbs, and views for each of the videos are listed in Table~\ref{tab:videostats}. Average duration of the videos is $77$ seconds.

\begin{table*}[tbh!]
\centering
\begin{tabular}{c|c|ccc|ccc|c|c}
\toprule
& & \multicolumn{3}{c|}{Comments} & \multicolumn{3}{c|}{Thumbs} & \\
  Survey &          Video &  Pos. &  Neut. &  Neg. &    Up &  Down & Fraction up &      Views &                                                        Link to the original \\
\midrule
         &       ski lift &   168 &    179 &    76 &    533 &    154 &      0.78 &    455,970 &  \href{http://youtu.be/GP2wvGVCsIU}{GP2wvGVCsIU} \\
         &            ufo &    55 &    136 &   317 &    140 &   1197 &      0.10 &    844,339 &  \href{http://youtu.be/PCMklx9YvHQ}{PCMklx9YvHQ} \\
         &    shark prank &    76 &     53 &    49 &    205 &    211 &      0.49 &    307,750 &  \href{http://youtu.be/Rk1LXZgCSpE}{Rk1LXZgCSpE} \\
  Part I &       fat talk &   296 &    203 &    75 &   2796 &     79 &      0.97 &    154,843 &  \href{http://youtu.be/V2SHwdtBH64}{V2SHwdtBH64} \\
         &     pony shoes &   196 &    217 &   653 &    439 &   1060 &      0.29 &    741,023 &  \href{http://youtu.be/hJJtVUTWCcc}{hJJtVUTWCcc} \\
         &  rollers trick &    24 &     41 &    35 &    638 &    164 &      0.80 &    254,765 &  \href{http://youtu.be/qgSv8B6UiUY}{qgSv8B6UiUY} \\
         &       cat bath &   148 &    156 &   105 &   1009 &    354 &      0.74 &    667,656 &  \href{http://youtu.be/xR6j4ECkDT4}{xR6j4ECkDT4} \\
\hline
         &   feeding croc &    48 &     16 &    34 &    576 &     58 &      0.91 &    836,942 &  \href{http://youtu.be/EPW0m0mc6hc}{EPW0m0mc6hc} \\
         &       veet add &    59 &     55 &   139 &    604 &   1377 &      0.30 &    661,311 &  \href{http://youtu.be/UxCHLXQffsg}{UxCHLXQffsg} \\
         &     google car &    89 &     26 &    68 &   1713 &     23 &      0.99 &    123,721 &  \href{http://youtu.be/aqrttLPjv1E}{aqrttLPjv1E} \\
 Part II &      baby yoga &    28 &     33 &   379 &    275 &   1996 &      0.12 &    870,166 &  \href{http://youtu.be/fFwrZHFLe2E}{fFwrZHFLe2E} \\
         &    all nighter &   288 &    582 &   232 &  12592 &    502 &      0.96 &  1,008,121 &  \href{http://youtu.be/kFcnUsYKT5w}{kFcnUsYKT5w} \\
         &        skywalk &    20 &     28 &    19 &    275 &     25 &      0.92 &    245,629 &  \href{http://youtu.be/laveE4bUm3M}{laveE4bUm3M} \\
         &     haribo add &    33 &     52 &    36 &    165 &     52 &      0.76 &    138,925 &  \href{http://youtu.be/qc8vxx6J5Xw}{qc8vxx6J5Xw} \\
\bottomrule
\end{tabular}

\caption{Basic statistics of the original, non-manipulated, videos: the number of comments of various types, the number of thumbs up and thumbs down, and the number of views at the time of their collection (February 2015).}
\label{tab:videostats}
\end{table*}

\subsection{Comments}

% labeling
Three labelers classified the comments as either positive, neutral, negative, or unreadable (see Table~\ref{tab:videostats} for a summary). The Fleiss' kappa between the three labelers is $0.56$. 
The comments that are unreadable or are written in a language different from English are filtered out. 
The ambivalent comments with conflicting labels, namely the comments that are labeled as both positive and negative by different labelers, are filtered out as well. In total, we filter out about $10\%$ of all comments. We apply a majority rule to determine the final label of each remaining comment. 
%We make the labeled comments available to scientific community.\footnote{The comments can be found at:\url{link}}. 

% tracking
During the experiment we measure which comments are shown to the participant on the screen. In order to see the comments, the participants need to scroll down, what allows precise measurements. Then, in the analysis, we estimate the number of comments read by the participant with the number of comments shown on the screen.
%, neglecting the fact that a given individual actually read the comments or just skimmed them. 
%That makes more difficult an analysis where the order of the comments is considered. However the fluctuations in the running average for a window of 50 points indicates that there are some effects on the order of the comments.

\subsection{Data Processing}

We apply the same filtering steps as for the data from Experiment I. As before, we discard the participants with incomplete answers, who gave answers more than once, or have answers invalidated for more than one videos (in total, less than $12\%$ of all participants).
%For the microscopic modeling, we additionally look into how many comments a participant read for each video. Since we have only seven measurements per each user, multiple measurements at $\Delta n=0$ make it harder to infer influenceability per each user. Thus, we invalidate the responses that have $\Delta n=0$. Finally, we discard the participants with answers invalidated for more than one videos (additional $8\%$ of all participants discarded).

\begin{figure}[htb]
\begin{picture}(290,135)
\centering
\put(0,0){
\includegraphics[width=0.24\textwidth]{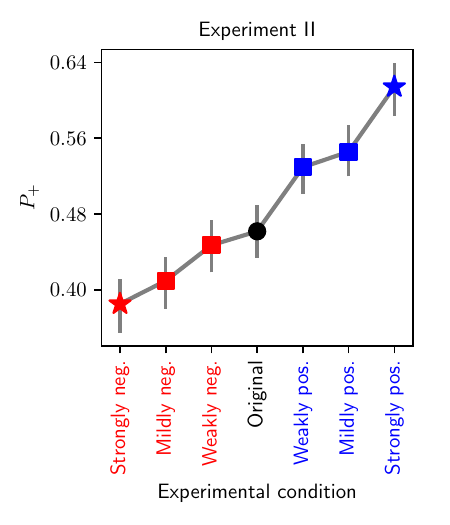}
\includegraphics[width=0.24\textwidth]{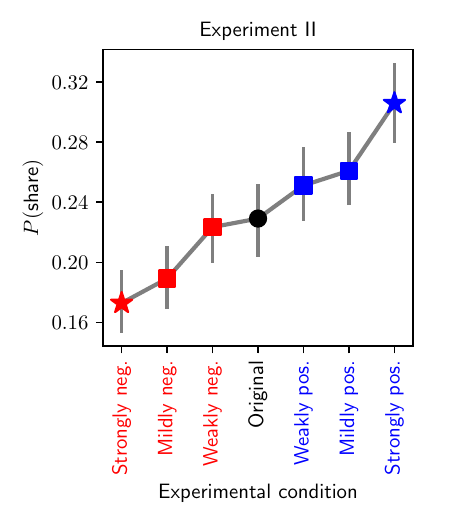}
}
\put(33,120){\small A}\put(160,120){\small B}
\end{picture}
\caption{The probability of positive opinion (left) and the sharing willingness (right) for the control condition (black circles) and the experimental conditions of various strength. The color of markers encodes positive (blue) and negative (red) conditions.}
\label{fig:exp2grad}
\end{figure}

\subsection{Demographics and Feedback of Participants}

\begin{figure}[tbh!]
\begin{picture}(290,190)
\centering
\put(0,100){
\subfigure{\includegraphics[width=0.24\textwidth]{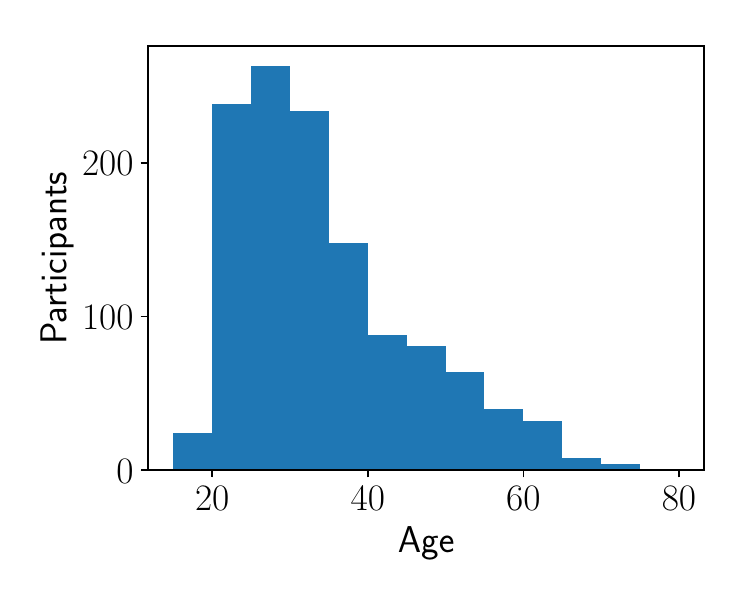}}
\subfigure{\includegraphics[width=0.24\textwidth]{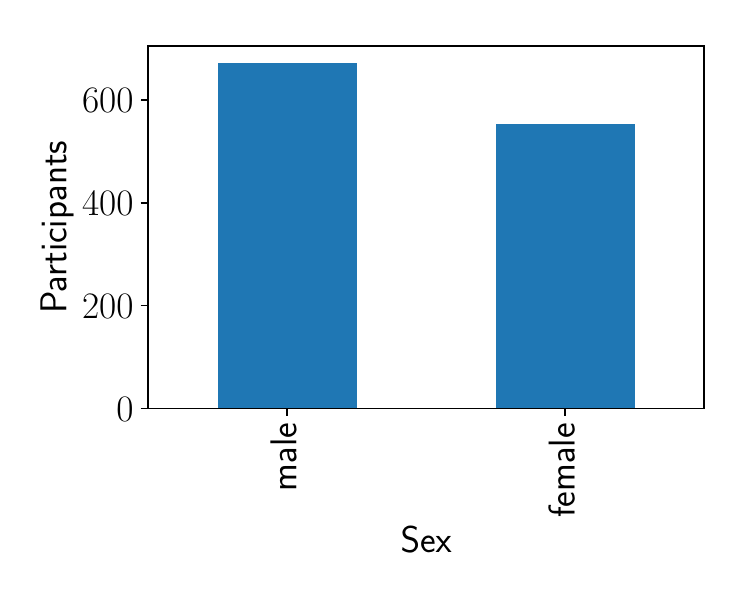}}
}
\put(0,0){
\subfigure{\includegraphics[width=0.24\textwidth]{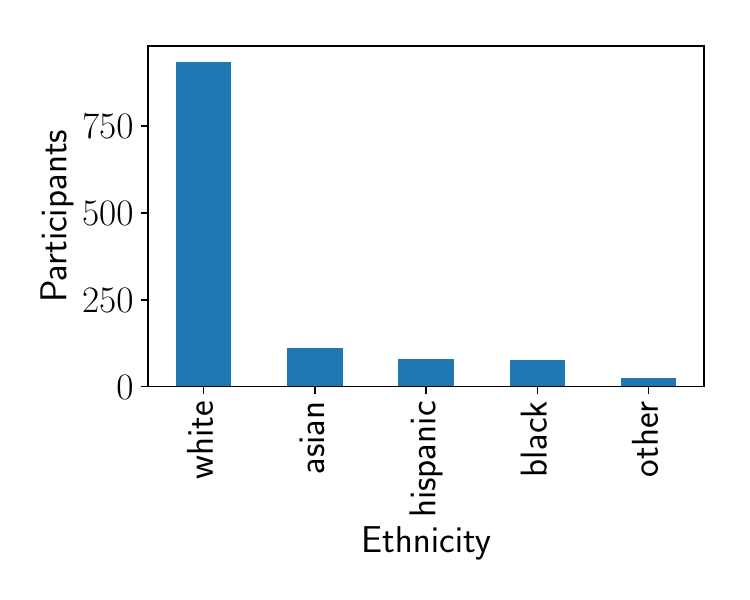}}
\subfigure{\includegraphics[width=0.24\textwidth]{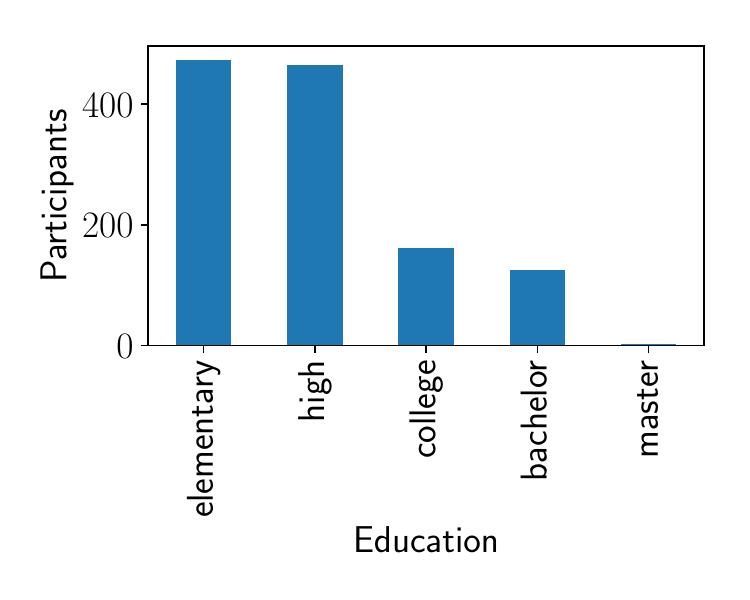}}
}
\put(110,182){\small A}\put(234,182){\small B}
\put(110,82){\small C}\put(234,82){\small D}
\end{picture}
\caption{Demographic information about the participant of the experiment.}
\label{fig:exp2demographics}
\end{figure}

The participants of our experiments are slightly more likely to be male than female, predominantly white, young adults, with varying levels of education (see Figure~\ref{fig:exp2demographics}).
Many of them enjoyed participating in our experiments. Most of the voluntary comments left by the participants after the experiments express their positive sentiment and gratitude, e.g.: ``I enjoyed this survey very much.'', ``great videos! expected something boring'', and ``fun videos. Thanks!''. Overall, $185$ participants voluntarily left positive feedback to our experiments containing at least one of these three words: ``enjoy'', ``great'', or ``fun''.

%\begin{figure}[hbtp]
%\centering
%\subfigure{\includegraphics[width=0.24\textwidth]{figs2/hist-age_survey3}}
%\subfigure{\includegraphics[width=0.24\textwidth]{figs2/hist-sex_survey3}}
%\subfigure{\includegraphics[width=0.24\textwidth]{figs2/hist-ethnicity_survey3}}
%\subfigure{\includegraphics[width=0.24\textwidth]{figs2/hist-education_survey3}}
%\caption{CAPTION}
%\label{fig:LABEL}
%\end{figure}

%``I enjoyed this survey very much.''
%``Really great''
%``great videos! expected something boring''
%``I really enjoyed this - it was fun to see some new things.''
%``fun videos. Thanks!''

%\section{Bayesian Social Influence}
%\label{sec:model-formulation}
%\input{sec_model-formulation}
%\input{sec_models-micro-mle}

\section{Models}
\label{sec:models-macro}

\subsection{Model Selection}

\begin{table*}[hbtp]
\centering
\begin{tabular}{l|l|c|c|c}
\toprule
Rank & Model & $-\log(P)$  & \#parameters & AIC \\
\midrule
1 &  $P(\matr {\Delta n}, \matr y | K=2 , \underset{K \times V}{\matr{p}} ,
\underset{1 \times V}{\matr{a_1}} , s_1 \equiv 0 ,
\underset{1 \times V}{\matr{a_2}} , s_2 )$
& 3316.4 &  43 & 6718.7 \\
2 &  $P(\matr {\Delta n}, \matr y | K=2 , \underset{K \times V}{\matr{p}} ,
a_1 , s_1 ,
\underset{1 \times V}{\matr{a_2}} , s_2 \equiv 0 )$
& 3330.7 &  30 & 6721.5 \\
3 &  $P(\matr {\Delta n}, \matr y | K=2 , \underset{K \times 1}{\matr{p}} ,
\underset{1 \times V}{\matr{a_1}} , s_1 \equiv 0 ,
\underset{1 \times V}{\matr{a_2}} , \underset{1 \times V}{\matr{s_2}} )$
& 3318.4 &  43 & 6722.8 \\
4 &  $P(\matr {\Delta n}, \matr y | K=2 , \underset{K \times V}{\matr{p}} ,
a_1 \equiv 0 , s_1 ,
\underset{1 \times V}{\matr{a_2}} , s_2 \equiv 0 )$
& 3332.4 &  29 & 6722.8 \\
5 &  $P(\matr {\Delta n}, \matr y | K=3 , \underset{K \times V}{\matr{p}} ,
a_1 , s_1 ,
a_2 , s_2 \equiv 0 ,
a_3 , s_3 \equiv 0 )$
& 3330.8 &  32 & 6725.5 \\
6 &  $P(\matr {\Delta n}, \matr y | K=2 , \underset{K \times V}{\matr{p}} ,
a_1 , \underset{1 \times V}{\matr{s_1}} ,
\underset{1 \times V}{\matr{a_2}} , s_2 \equiv 0 )$
& 3320.1 &  43 & 6726.2 \\
7 &  $P(\matr {\Delta n}, \matr y | K=3 , \underset{K \times V}{\matr{p}} ,
a_1 , s_1 \equiv 0 ,
a_2 , s_2 \equiv 0 ,
a_3 \equiv 0 , s_3 )$
& 3332.4 &  31 & 6726.8 \\
8 &  $P(\matr {\Delta n}, \matr y | K=3 , \underset{K \times 1}{\matr{p}} ,
\underset{1 \times V}{\matr{a_1}} , \underset{1 \times V}{\matr{s_1}} ,
\underset{1 \times V}{\matr{a_2}} , s_2 \equiv 0 ,
a_3 \equiv 0 , \underset{1 \times V}{\matr{s_3}} )$
& 3307.5 &  58 & 6730.9 \\
9 &  $P(\matr {\Delta n}, \matr y | K=3 , \underset{K \times 1}{\matr{p}} ,
\underset{1 \times V}{\matr{a_1}} , \underset{1 \times V}{\matr{s_1}} ,
\underset{1 \times V}{\matr{a_2}} , s_2 \equiv 0 ,
a_3 \equiv 0 , s_3 \equiv 0 )$
& 3322.9 &  44 & 6733.8 \\
10 &  $P(\matr {\Delta n}, \matr y | K=2 , \underset{K \times V}{\matr{p}} ,
a_1 \equiv 0 , \underset{1 \times V}{\matr{s_1}} ,
\underset{1 \times V}{\matr{a_2}} , s_2 \equiv 0 )$
& 3325.0 &  42 & 6733.9 \\
\bottomrule
\end{tabular} 
\caption{The list of top variants of the sub-population model. The variants are ranked by the value of AIC. All top variants have $s_k \equiv 0$ for at least one sub-population.}
\label{tab:variants}
\end{table*}

\begin{table*}[htb]
\centering
\begin{tabular}{c|c|c|ccc|c|ccc|c|ccc}
% & \multicolumn{2}{c}{actual class (observation)} \\ 
%\hline 
%\multicolumn{2}{c}{predicted class (expectation)} 
% & tp (correct result) & fp (unexpected result) \\ 
% & fn (missing result) & tn (correct absence of result)
\toprule
Survey & Video & $p_2$ & \multicolumn{3}{c|}{$\sigma(p_2)$}  & $a_2$ & \multicolumn{3}{c|}{$\sigma(a_2)$}  & $a_1$ & \multicolumn{3}{c}{$\sigma(a_1)$} \\
\midrule
         &       ski lift &  0.32 &  0.01 &  0.05 &  0.06 &   3.44 &  2.29 &  5.90 &  1.28 & -0.75 &  0.02 &  0.23 &  0.20 \\
         &            ufo &  0.11 &  0.02 &  0.06 &  0.04 &  -1.87 &  3.35 &  5.37 &  2.43 & -3.84 &  0.08 &  0.40 &  0.38 \\
         &    shark prank &  0.16 &  0.01 &  0.05 &  0.05 &   1.77 &  0.60 &  5.01 &  1.70 & -1.47 &  0.02 &  0.19 &  0.21 \\
  Part I &       fat talk &  0.17 &  0.01 &  0.07 &  0.06 &  -8.28 &  3.19 &  8.95 &  2.00 &  0.80 &  0.07 &  0.30 &  0.19 \\
         &     pony shoes &  0.12 &  0.00 &  0.10 &  0.07 &  -9.67 &  4.07 &  7.01 &  3.99 & -2.37 &  0.02 &  0.22 &  0.21 \\
         &  rollers trick &  0.40 &  0.01 &  0.04 &  0.05 &  -0.22 &  0.13 &  1.16 &  0.74 & -0.25 &  0.01 &  0.17 &  0.17 \\
         &       cat bath &  0.35 &  0.01 &  0.05 &  0.05 &   1.39 &  0.36 &  2.89 &  0.86 &  0.17 &  0.01 &  0.17 &  0.17 \\
\midrule
         &   feeding croc &  0.29 &  0.07 &  0.16 &  0.06 &  15.51 &  1.58 &  3.74 &  1.28 &  0.59 &  0.16 &  0.43 &  0.20 \\
         &       veet add &  0.15 &  0.03 &  0.12 &  0.04 &   3.04 &  0.65 &  2.32 &  2.43 & -0.15 &  0.03 &  0.68 &  0.38 \\
         &     google car &  0.05 &  0.01 &  0.13 &  0.05 &   1.59 &  1.59 &  3.64 &  1.70 &  2.43 &  0.05 &  0.39 &  0.21 \\
 Part II &      baby yoga &  0.20 &  0.05 &  0.16 &  0.06 &   0.50 &  0.81 &  2.00 &  2.00 & -2.52 &  0.08 &  0.42 &  0.19 \\
         &    all nighter &  0.15 &  0.03 &  0.10 &  0.07 &  -3.60 &  1.68 &  1.96 &  3.99 &  1.06 &  0.07 &  0.63 &  0.21 \\
         &        skywalk &  0.18 &  0.05 &  0.18 &  0.05 &   1.47 &  0.31 &  1.99 &  0.74 &  1.89 &  0.05 &  1.07 &  0.17 \\
         &     haribo add &  0.21 &  0.04 &  0.12 &  0.05 &  -0.93 &  0.47 &  1.55 &  0.86 &  0.89 &  0.04 &  0.31 &  0.17 \\
\bottomrule
\end{tabular} 
\caption{The values of parameters of the best sub-population model and their standard error. The standard error is estimated using three methods, from left to right: jackknife re-sampling, bootstrap re-sampling, and observed Fisher information. Note that the standard error of prior opinions of influenceable sub-population is many times larger, i.e., $\sigma(a_2) > \sigma(a_1)$. 
}
\label{tab:bestparams}
\end{table*}

The model of sub-populations does not determine what is the best number of sub-populations nor whether all parameters $p_{kv}$, $a_{kv}$, and $s_{kv}$ are indeed necessary. Each of these parameters can depend on the video, can be shared across the videos, e.g., 
$p_{kv} \equiv p_{k}$, 
or vanish by being replaced with a fixed neutral value across videos, i.e., 
$p_{kv} \equiv 1/K$, or
$\displaystyle\mathop{\forall}_{v} a_{kv} = 0$, or
$\displaystyle\mathop{\forall}_{v} s_{kv} = 0$.
We represent the structure of the model using the aforementioned notation for the likelihood,
\begin{equation}
P(\matr {\Delta n}, \matr y | k, \underset{K \times V}{\matr{p}}, \underset{1 \times V}{\matr{a_1}}, \underset{1 \times V}{\matr{s_1}}, \dots, \underset{1 \times V}{\matr{a_K}}, \underset{1 \times V}{\matr{s_K}}  ),
\end{equation}
which states that the parameters $p$, $a_1$, $s_1$, $\dots$, $a_K$, $s_K$ depend on the videos. As an example, we show the representation of the model with $K=2$ sub-populations, vanishing $s_1$, parameter $s_2$ independent from videos, and the remaining parameters dependent on videos:
\begin{equation}
P(\matr {\Delta n}, \matr y | K=2 , \underset{K \times V}{\matr{p}} ,
\underset{1 \times V}{\matr{a_1}} , s_1 \equiv 0 ,
\underset{1 \times V}{\matr{a_2}} , s_2 ),
\label{eq:best}
\end{equation}
which is the variant introduced in the main text of this manuscript that maximizes Akaike information criterion (AIC). 
In other words, each of the parameters of a model variant takes one of $N=3$ different forms and each sub-population is defined by one of $N^2$ combinations of parameters. The sub-populations are interchangeable, so in total there is $N \times \left[ \binom{N^2+K-1}{K} \right]$ variants of the model for $K$ sub-populations. Note, however, that a model with more than one sub-population with vanishing parameters, i.e., multiple sub-populations $k$ such that 
$\displaystyle\mathop{\forall}_{v} a_{kv} = 0$ and  $\displaystyle\mathop{\forall}_{v} s_{kv} = 0$, 
is equivalent to a model having just one such sub-population. Hence, overall, for $K$ sub-populations there are $N \times \left[ \binom{N^2+K-1}{K} - \binom{N^2+K-3}{K-2} \right]$ different variants of the sub-populations model. 

\begin{figure}[hbt]
\centering
\includegraphics[width=0.44\textwidth]{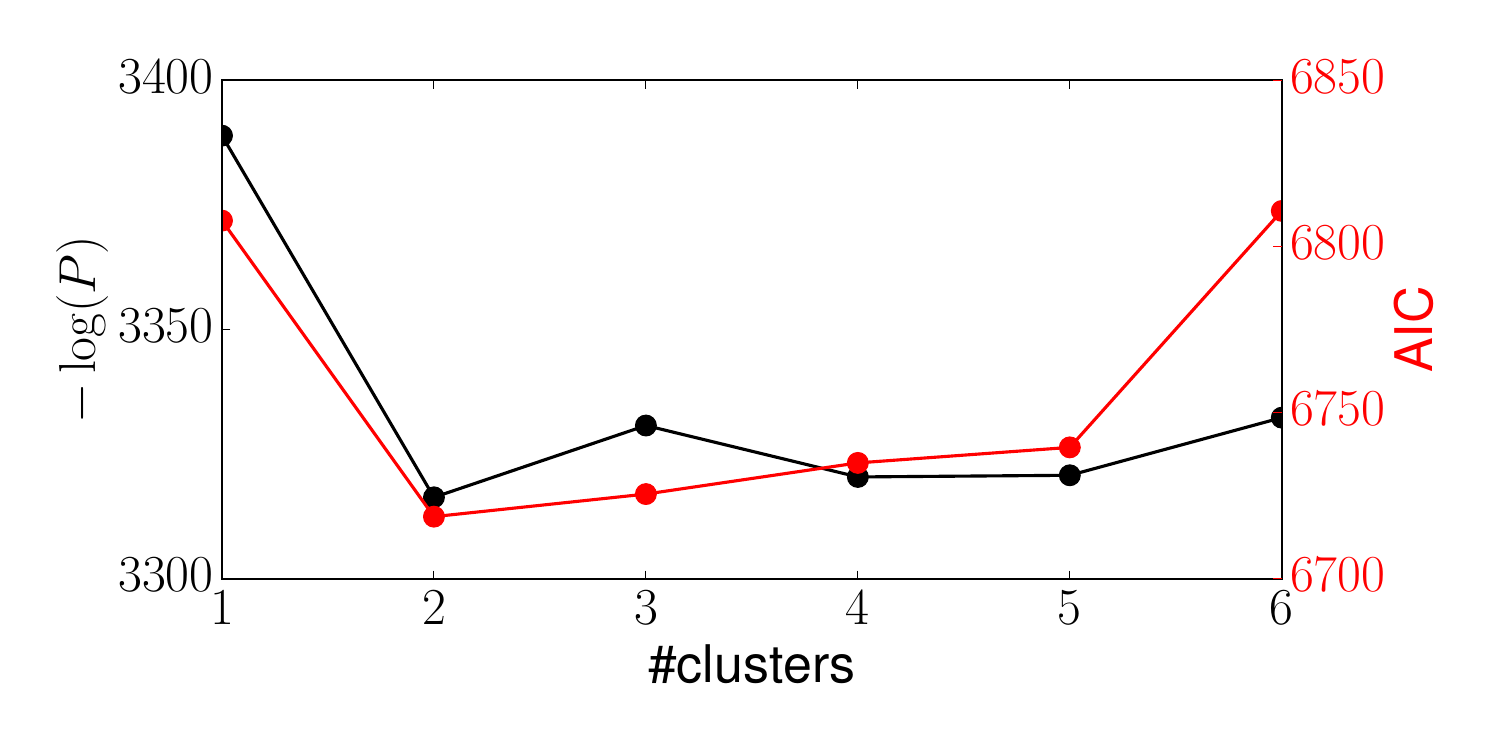}
\includegraphics[width=0.44\textwidth]{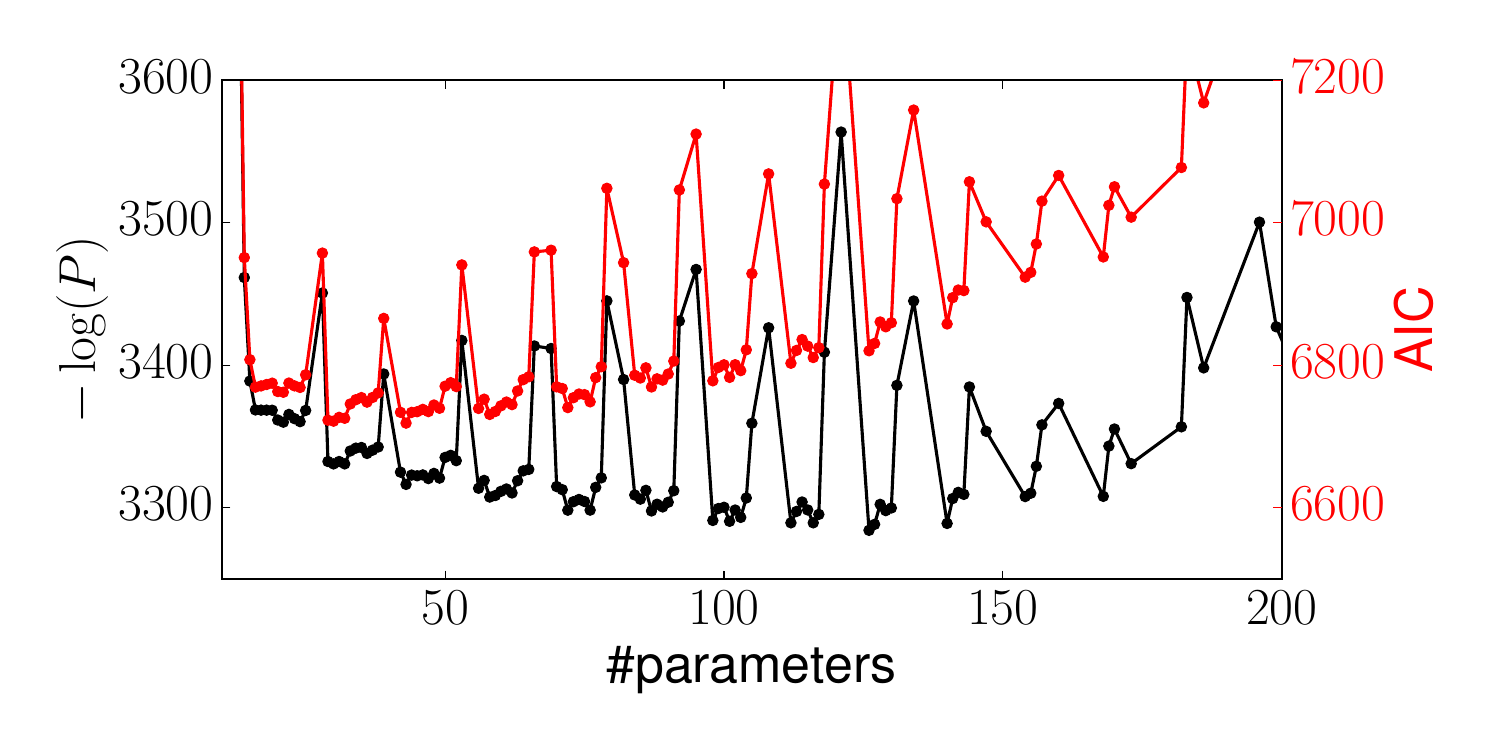}
\caption{The minimal negative log-likelihood and Akaike information criterion of the sub-population model having a given number of clusters (top) or parameters (bottom).}
\label{fig:variants}
\end{figure}

We fit each variant of sub-populations model for $K \in \{1,2,\dots,6\}$ by maximizing the log-likelihood of the corresponding model (Equation \ref{eq:p-complete}). 
To this end, we perform $1000$ realizations of the L-BFGS-B algorithm, each time initializing the parameters with random values sampled from a standard normal distribution, with the exception of $p_{kv}$, which is always initialized with the uniform distribution of individuals over sub-populations, i.e., $p_{kv}=1/K$.
% results
Different variants of sub-populations model generally have varying number of parameters and sub-populations. After fitting these variants of the model, we compare them using AIC, which penalizes for the number of parameters. We plot the best values of AIC and log-likelihood against the number of parameters and sub-populations (Figure~\ref{fig:variants}). The models achieving low AIC tend to have two sub-populations and less than $50$ parameters, that is about $3$ parameters per video.
% best
The top ten best variants of the models are shown in Table~\ref{tab:variants}. Out of these ten models, six have two sub-populations, while four have three sub-populations. In all cases, among these sub-populations, there is at least one non-influenceable sub-population ($\displaystyle\mathop{\forall}_{v} s_{kv} = 0$). The four top variants of the sub-population model are very similar to the top model presented in the main text, while the remaining top models are also similar, but may have an additional sub-population. Overall, the findings shown in the main text for the top model are consistent with the other top ten models.

\subsection{Parameters of the Best Model}

\begin{figure}[htb]
\centering
\includegraphics[width=0.5\textwidth]{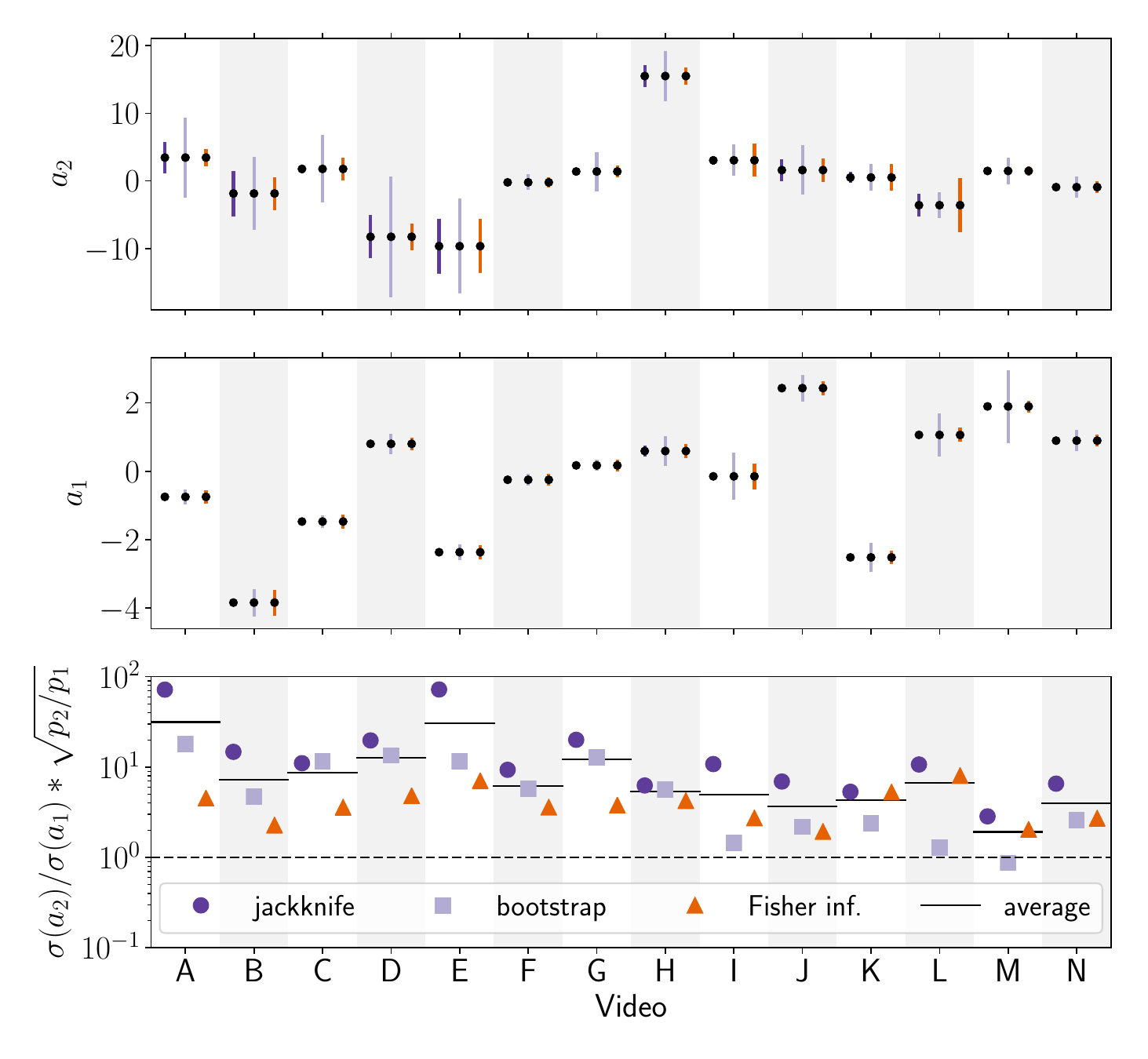}
\caption{The values of parameters with their standard errors of the best sub-population model for influenceable (top) and non-influenceable (middle) sub-population. The standard deviation is computed using three different methods, from left to right: jackknife re-sampling, bootstrap re-sampling, and the inverse of observed Fisher information. The bottom figure shows the ratio of sub-populations' standard-deviation corrected for their sizes.}
\label{fig:bestparams}
\end{figure}

The best sub-population model is 
$P(\matr {\Delta n}, \matr y | K=2 , \underset{K \times V}{\matr{p}} ,
\underset{1 \times V}{\matr{a_1}} , s_1 \equiv 0 ,
\underset{1 \times V}{\matr{a_2}} , s_2 )$. 
It has two sub-populations: non-influenceable sub-population with influenceability $s_1 \equiv 0$ and influenceable sub-population having $s_2 = 0.8 \pm 0.005$. 
% findings
We list the value of the remaining parameters for each video in Table \ref{tab:bestparams}. 
Individuals in influenceable sub-population have much noisier prior opinions about each video than individuals in non-influenceable sub-populations (Figure~\ref{fig:bestparams}).

\pnasbreak

\end{document}